\documentclass[preprint,prb,superscriptaddress,amsfonts,amsmath,amssymb,showpacs]{revtex4}
\allowdisplaybreaks[1]
\usepackage{graphicx}
\newcommand{\rem}[1]{}
\newcommand{\ket}[1]{\left| #1 \right\rangle}
\newcommand{\bra}[1]{\left\langle #1 \right|}
\newcommand{\mC}{\mathcal{C}}
\newcommand{\mD}{\mathcal{D}}
\newcommand{\mE}{\mathcal{E}}
\newcommand{\mF}{\mathcal{F}}
\newcommand{\be}{\begin{equation}}
\newcommand{\ee}{\end{equation}}
\begin{document}
\title{
Robust non-adiabatic molecular dynamics for metals and insulators
}
\author{L. Stella}
\email{l.stella@ucl.ac.uk}
\affiliation{Department of Physics and Astronomy, University College
  London, Gower Street, London WC1E 6BT, UK}
\affiliation{London Centre for Nanotechnology, 17--19 Gordon Street,
  London WC1H OAH, UK}
\author{M. Meister}
\affiliation{Department of Physics and Astronomy, Queen's University
  Belfast, Belfast BT7 1NN, UK}
\author{A.J. Fisher}
\affiliation{Department of Physics and Astronomy, University College
  London, Gower Street, London WC1E 6BT, UK}
\affiliation{London Centre for Nanotechnology, 17--19 Gordon Street,
  London WC1H OAH, UK}
\author{A.P. Horsfield}
\affiliation{Department of Materials, Imperial College London, South
  Kensington Campus, London SW7 2AZ, UK}
\begin{abstract}
We present a new formulation of the correlated electron-ion
dynamics (CEID) scheme, which systematically improves Ehrenfest
dynamics by including quantum fluctuations around the mean-field
atomic trajectories. 
We show that the method can simulate models of
non-adiabatic electronic transitions, and test it against exact
integration of the time-dependent Schr\"{o}dinger equation.
Unlike previous formulations of CEID, the accuracy of this scheme
depends on a single tunable
parameter which sets the level of atomic fluctuations included.
The convergence to the exact dynamics
by increasing the tunable parameter is demonstrated for
a model two level system.
This algorithm provides a smooth description of the non-adiabatic
electronic transitions which satisfies the kinematic
constraints (energy and momentum conservation)
and preserves quantum coherence.
The applicability of this algorithm to more complex atomic
systems is discussed.
\end{abstract}
\maketitle
%
\section{Introduction}\label{Intro:sec}
%
Ordinary molecular dynamics (MD) 
\cite{tuckerman00} 
uses
Hamiltonian equations of motion (EOM) to describe the evolution of an
atomic system.
The validity of that approach relies on the Born-Oppenheimer (BO)
approximation 
which states that --- in most cases ---
the atomic motion induces  
a slow (adiabatic) perturbation of the electronic dynamics.
Therefore, if the system is originally prepared in an electronic
eigenstate (e.g. the ground-state)
for a given atomic configuration, 
it will evolve into the same electronic eigenstate
for the evolved atomic configuration.
Furthermore, according to the BO approximation, the ground-state
electronic energy for a given atomic
configuration can be employed as 
an effective (i.e. low energy) \emph{potential energy surface} (PES) 
for the atomic motion.

The BO approximation is not longer applicable whenever
electronic transitions between surfaces are relevant: 
for instance, where there is a non-radiative transition following
an earlier photo-excitation.
However, it is still possible to extend the scope of MD by allowing
more than a single PES --- one for every electronic state --- and by
providing a meaningful way to make transitions among PES i.e.
\emph{non-adiabatic electronic transitions}. 

The extension of ordinary MD to deal with processes 
involving many PES has been extensively pursued over recent
decades and some effective algorithms 
are available for atomistic simulations.
The most used are:
Ehrenfest dynamics (ED),
\cite{tully98,horsfield06a} 
molecular dynamics with quantum transitions (MDQT),
\cite{hammesschiffer94,tully98},
mixed quantum-classical dynamics (MQCD),
\cite{kapral99a,kapral06} 
and \emph{ab initio} multiple spawning (AIMS) 
\cite{ben-nun00,ben-nun02}.
They have been employed to simulate 
quantum dissipative dynamics
\cite{mackernan02,sergi07a}
as well as
non-adiabatic processes
such as photo-chemical reactions,
\cite{mueller97,kin07}
polaron formation in conjugated polymers,
\cite{an04},
and proton transfer in solutions 
\cite{hammesschiffer94}. 
Although less successful in practical applications, we also mention
other dynamical schemes that provide valuable theoretical insights:
the frozen Gaussian approximation,
\cite{heller81}
which is one of the pillars of AIMS, and the Gauss-Hermite
wave-packet expansion,
\cite{adhikari99}
which shares similarities to the method presented in this paper.

In order to extend the scope and the accuracy
of the aforementioned methods, a new approach called 
\emph{correlated electron-ion  dynamics} (CEID)
\cite{horsfield04b,horsfield05a,horsfield06a}, 
has been introduced.
This method has been mainly applied
to study the heat production and dissipation in model 
metallic nanostructures,
\cite{bowler05a,horsfield06a,mceniry07}
a problem relevant for nanotechnology.
Indeed, other algorithms based on non-smooth dynamics, that is, those 
which allow for either sudden surface hopping (like MDQT)
or sudden wave-packet spawning (like AIMS), are expected 
to be less efficient in the simulation of systems with a dense, gapless
electronic spectrum, like a metal.
That is a consequence of the slight adjustments of the average
atomic positions and/or velocities that might need to be imposed 
after a sudden
transition in order to ensure total energy and momentum are conserved.
\cite{tully98,ben-nun02}
If the number of crossings is large, the computational overhead due to
these adjustments might be non-negligible.
MQCD, although it is often implemented by using both
surface hopping and mean-field evolutions, is based on a sound theory
\cite{nielsen01,sergi05}
and conserves the kinematic constraints.
On the other hand, time-translation invariance is valid only
approximately in MQCD
\cite{nielsen01}
and numerical instabilities have so far limited the surface-hopping
implementation of MQCD to relatively short time simulations.
\cite{sergi07a}
Finally, ED --- although it evolves smoothly --- 
has been shown to poorly describe the atomic heating caused by
electron-ion interaction.
\cite{horsfield04a} 

A further complication arises when a quantum sub-system is coupled
with large quantum reservoirs (which are not treated in detail),
as for a nanostructure connected to macroscopic leads.
In this case, it is hard to define meaningful PES in terms of
the sub-system degrees of freedom only, and so algorithms based on the
surface hopping paradigm are expected to be less accurate.
This problem is absent if the quantum sub-system is coupled to a
classical dissipative environment; in this case  MQCD or its latest
variant 
\cite{sergi07b}
can give reliable results.

In principle, the CEID EOM form an exact, yet infinite,
kinetic hierarchy which corrects ED by means of the so-called
small amplitude moment expansion (SAME) of the Liouville equation.
\cite{horsfield06a}
So far, a few schemes to truncate the hierarchy have been proposed
\cite{horsfield04b} 
in order to simulate the CEID EOM and 
currently available algorithms are
restricted to a mean-field second moment approximation.
\cite{footnote4,horsfield05a}
Although those algorithms are accurate enough to describe --- 
at least qualitatively --- nanostructure heating, 
the existence of a practical truncation scheme which converges
to the exact quantum dynamics has not been demonstrated until now:
in this paper we show that a convergent truncation scheme actually
exists and we provide a new practical CEID algorithm whose accuracy
depends on a single tunable parameter.
We provide a validation of our theory 
by comparing  CEID results against exact integration of the
time-dependent Schr\"{o}dinger equation for 
a model two level
system (2LS) --- i.e. the simplest system
which displays non-adiabatic transitions
\cite{landau32b,zener32}
--- in two contrasting parameter regimes.

The rest of the paper is organized as follows:
In Sec.~\ref{2LS:sec} we introduce the physics of the model 2LS employed to
test our new CEID algorithm.
In Sec.~\ref{exact_int:sec} we describe the exact algorithm by which
we produced the
benchmark calculations, while in Sec.~\ref{CEID:sec} the new
CEID scheme is derived in detail.
Finally, numerical results from simulations of the model 2LS
are collected in Sec.~\ref{2LS_results:sec} and the
conclusions and perspectives 
of this work are discussed in Sec.~\ref{conclusions:sec}. 
%
\section{A case study: the two level system}\label{2LS:sec}
%
In this section we introduce the model 2LS
that we employed to test the convergence properties of our CEID scheme. 
[Numerical findings are reported in Sec.~\ref{2LS_results:sec}.]
Here we discuss the physics we expect to address
before explaining the details of the algorithms we use.

It is widely recognized that a one-dimensional 
2LS illustrates many of the
fundamental features of a non-adiabatic system and, at the same time,
it retains the simplicity of a low-dimensional model.
\cite{tully90,martens97}
In general, the Hamiltonian for a system made by electrons and ions
(in the absence of external fields) can be written as follows:
\begin{equation}\label{2LS_H:eqn}
\hat{H} =  \hat{P}^2/2M + \hat{H}_{e}(\hat{R})\;,
\end{equation}
where $\hat{R}$ and $\hat{P}$ are
the quantum operators for the atomic position and
momentum, while the electronic dependence of $H$ is collected
into $\hat{H}_{e}$.
In particular, the first term in the RHS of
Eq.~\eqref{2LS_H:eqn} accounts for the atomic kinetic operator while
a sort of atomic potential operator is described by the second
term.

There is a lot of freedom in constructing the
potential term, $\hat{H}_{e}$, but
in this paper we focus only on the following parametrization:
\begin{equation}\label{elec_H:eqn}
\hat{H}_{e}(R) = \left(
\begin{array}{cc}
\frac{1}{2}K\,\left( R -R_{0}\right)^2 & -f_c \,R\\
-f_c \,R  & \frac{1}{2}K\,R^2 + \Delta\varepsilon
\end{array}
\right)\;
\end{equation}
which describes two \emph{parabolic} PES (diagonal
entries) linearly coupled through a kind of
dipolar interaction (off-diagonal entries).
This is a \emph{non-adiabatic} representation of the electronic PES in which
the electronic basis is independent of the atomic coordinate.
The \emph{adiabatic} representation can be
obtained as usual by diagonalizing $\hat{H}_{e}(R)$.
Eq.~\eqref{2LS_H:eqn} 
and \eqref{elec_H:eqn} depend on the following parameters:
the atomic mass $M$, the harmonic constant $K$, the electron-ion 
coupling constant $f_c$, 
the surface displacement $R_0$, and the PES energetic offset
$\Delta \varepsilon$.

Since both the PES are confining,
we expect to see periodic electronic transitions between
the two PES driven by the electron-ion interaction. 
For instance, a state can be prepared as the atomic ground-state on the
upper electronic PES.
The atomic position $R$ experiences quantum oscillations and so the
system cannot be exactly localized in
the minimum of the PES ($R=0$ according to our parametrization).
Since this initial
state is not an eigenstate of the interacting Hamiltonian (i.e. for
$f_c \neq 0$), 
the system  will eventually make a transition into the lower PES.
We stress that this process must conserve the total energy so that an
atomic transition must accompany the electronic transition.
For instance, the electronic process described above 
can be viewed as an initial decay
since the atomic potential energy is effectively decreased. 
Therefore, an increase of the atomic kinetic energy is expected as a
consequence of this decay in order to conserve the total energy.
This, in a nutshell, is the heating of an atomic degree of freedom
caused by the electron-ion interaction.
\cite{horsfield04a}
However, we stress that the aim of the present work is not
the study of a quantum decay process, but the illustration of the
convergence properties of a new CEID algorithm.
Other models must be used to address the physics of quantum
thermalization.
For instance, the spin-boson model
\cite{leggett87}
describes a 2LS coupled
to an environment made by a collections of many quantum
harmonic oscillators.
This model can be effectively simulated
\cite{makri98,mackernan02}
and it might provide a future test case 
for our new CEID algorithm.

There is an interesting general feature displayed by
the electronic dynamics of our model 2LS.
According to the initial condition described above, the electronic
transition is due to the quantum fluctuations only, because the dipolar
interaction is exactly zero for a classical
atom perfectly localized in the minimum of the upper PES.
On the other hand, quantum fluctuations are completely neglected in
the sort of mean-field description of the atomic motion employed in
ED.
As a consequence, we do not expect ED to reproduce the initial
transition from the upper PES to the lower PES,
which can be thought of as spontaneous phonon emission.
If confirmed, this behavior will provide further evidence that 
the exchange of energy from the electronic to the atomic degrees of
freedom cannot
be properly addressed by ED.
\cite{horsfield04a}

In order to be as transparent as possible when discussing the dynamical features
of our model 2LS,
we choose to measure the values of the parameters in the Hamiltonian 
in terms of natural units.
The natural energy scale is given by the harmonic quantum,
$\hbar\omega$, where $\omega = \sqrt{K/M}$, and so the time can be
measured in units of the harmonic period, $2\pi/\omega$.
Introducing a mass scale, $M_0$, (its
actual value is not important here), 
a length scale and a linear momentum scale are immediately obtained:
$a_0= \sqrt{\hbar/(M_0\omega)}$ and $b_o=\sqrt{M_0\hbar\omega}$,
respectively. 

For our purposes, not all the parameters need to be varied during the
numerical experiments.
First of all, we fixed both the atomic mass ($M=M_0$) and the harmonic
constant ($K=M_0\omega^2$) and then  
we took the PES offset to be equal to one
harmonic quantum ($\Delta \varepsilon = \hbar\omega$).
This is equivalent to saying that the atomic ground-state on
the upper PES, $|\chi_0^{(u)}\rangle$, 
has got exactly the same
energy as the first harmonic excitation on
the lower PES, $|\chi_1^{(l)}\rangle$.
[Here we are neglecting the electron-ion coupling and
using the notation introduced in appendix \ref{time_dep_pert:sec}.]
If the coupling constant is small enough
(see appendix \ref{time_dep_pert:sec}), the time-evolution of
our 2LS can be understood starting from these two states.
We confined this study to this weak coupling regime,
i.e. $f_c \le 0.1 \, \hbar \omega /a_0$.
Finally, we report in this paper the numerical results 
for two different 2LS geometries only:
the \emph{unshifted} 2LS, with $R_0=0$, 
and the \emph{shifted} 2LS, with $R_0=a_0$.
The PES for these two cases are shown in
Fig.~\ref{fig_potential_populations_unshifted:fig}(a) and
Fig.~\ref{fig_potential_populations_shifted:fig}(a), respectively.
Although many other 2LS geometries have been studied, e.g. with
larger $R_0$ or 
with other values of $K$ 
(including the possibility of different harmonic force constants for the
lower and the upper PES),
they gave numerical outcomes qualitatively similar to either the shifted or
unshifted 2LS, and so the details are not reported here.

A linear combination of the two
low-lying resonant states might be employed to give 
a qualitative account of 
our model 2LS dynamics i.e. 
the wave-function at time $t$ might be approximated as:
\begin{equation}\label{0-order_state:eqn}
\psi(t) \simeq c_0(t)|\chi_0^{(u)}\rangle
+ c_1(t)|\chi_1^{(l)}\rangle\;,
\end{equation}
where $c_0$ and $c_1$ are time-dependent complex coefficients.
This state would give a distribution of the atomic
position $R$ more or less localized around each of the two classical equilibrium
positions, namely $R=0$ for the upper PES and $R=R_0$ for the lower PES.
Eq.~\eqref{0-order_state:eqn} 
might not be a good \emph{ansatz} for
the evolved state as soon as 
the displacement $R_0$ is large enough.
Indeed, even a classical atom
can be found far from its equilibrium position whenever this is
energetically allowed.
In this case, we should be able to describe an atomic wave-packet
almost localized far from both $R=0$ and $R=R_0$.
Therefore, we will produce a manifest physical inconsistency 
if we assume that the generic 2LS state can be
always well approximated by Eq.~\eqref{0-order_state:eqn}.
This physical inconsistency can be easily fixed by taking a longer
expansion
of the exact evolved state $\psi(t)$ 
in terms of the harmonic excitations of the
atomic degrees of freedom.
Unfortunately, there is a cost to pay: the longer --- and so the more
accurate --- the expansion, 
the more time-consuming will be the simulation.
[See Sec.~\ref{conclusions:sec} for a further discussion of this
point.]

In the next two sections, two different ways to compute
non-adiabatic dynamics are considered in detail.
In Sec.~\ref{exact_int:sec} a method based on the numerical
diagonalization of the Hamiltonian of the full system is
presented,
while a new formulation
of CEID is introduced in Sec.~\ref{CEID:sec}.
%
\section{Exact integration}\label{exact_int:sec}
%
The `exact' method of integration is based on numerical
diagonalization of the full Hamiltonian matrix and subsequent time
evolution exploiting the eigenvectors and eigenvalues 
obtained in the diagonalization step. While it is highly accurate,
in practice, such an approach naturally has to be 
limited to systems with a small number of degrees of freedom. 
However, for our purposes it provides an ideal scheme for producing benchmark results. 
A central quantity in this paper is the Wigner transform (WT) of an
operator. 
Originally, the
WT of a wave-function $\psi$ was defined in
Ref.~\onlinecite{wigner32} as
\begin{align}
\label{MatRWig1}
W(R,P) &=& \frac{1}{{(\hbar\pi)}^{n}}\int\psi^{*}(R+s)\psi(R-s)
\exp\Big(\frac{2i}{\hbar}P\cdot s\Big){\text{d}}^{n}s \nonumber \\
&=&\frac{1}{{(2\pi\hbar)}^{n}}\int
\psi^{*}\big(R+\frac{1}{2}s\big)\psi(R-\frac{1}{2}s\big)\exp\Big(\frac{i}{\hbar}P\cdot s\Big)
{\text{d}}^{n}s,
\end{align}
where $R=(R_{1},\dots,R_{n})$, $P=(P_{1},\dots,P_{n})$, and the time-dependence
has been suppressed. 
The latter form in Eq.~\eqref{MatRWig1} is more common
nowadays.
In a straightforward extension of this definition,
for an operator $\hat{A}$, expandable in basis states $\{\phi_{a}\}$, 
$\hat{A}=\sum_{ab}\phi_{a}(x)A^{ab}\phi^{*}_{b}(x)$, the WT is given by
\begin{multline}
A_{w}(R,P)=\sum_{ab}A^{ab}\frac{1}{{(2\pi\hbar)}^{n}}
\int\phi^{*}_{b}\big(R+\frac{1}{2}s\big)\phi_{a}\big(R-\frac{1}{2}s\big)
\exp\Big(\frac{i}{\hbar}P\cdot s\Big){\text{d}}^{n}s.
\end{multline} 
Of particular interest to us here, however, are WT with
respect to the degrees of freedom of a 
subsystem, i.e. partial WT. 
These appear naturally when the system can be divided into subsystems
on physical grounds. 
In this paper, where the system is a
molecule, the system can be split into an electronic
subsystem and the subsystem of the ions/nuclei. 
Consider an operator $\hat{C}$ acting on a system divisible
into subsystems with  basis sets
$\{\ket{A}\}$ and $\{\ket{a}\}$. 
The operator is assumed to be expandable as
$\hat{C}=\sum_{ABab}\ket{a}\otimes\ket{A}C^{ABab}\bra{b}\otimes\bra{B}$,
and its partial WT with respect to the
$\ket{A}$-subsystem is 
\begin{equation}
\hat{C}_{w,A}(R,P) = \sum_{ab}\ket{a}C^{ab}_{w,A}(R,P)\bra{b},
\end{equation}
where, with $\Phi_{A}$ the position representation of $\ket{A}$, 
\begin{multline}
C^{ab}_{w,A}(R,P)=\sum_{AB}C^{ABab}\frac{1}{{(2\pi\hbar)}^{n}}
\int\exp\Big(\frac{i}{\hbar}P\cdot s\Big)
\Phi_{B}^{*}\Big(R+\frac{1}{2}s\Big)\Phi_{A}\Big(R-\frac{1}{2}s\Big){\text{d}}^{n}s.
\end{multline} 
Note the important fact that $\hat{C}_{w,A}$ is still an operator in
the $\ket{a}$-subsystem.  
As in the rest of the paper the only WT used are
partial WT with respect to `ionic' or `atomic' 
degrees of freedom, we shall henceforth write $\hat{C}_{w}$ instead of
$\hat{C}_{w,A}$, and also refer to the
$\ket{A}$-subsystem as the atomic and the $\ket{a}$-subsystem as the
electronic subsystems.

The time-evolution of the system is generated by a Hamiltonian, whose
eigenvalues and eigenvectors shall be
$\mE_{n}$ and $\ket{\Psi_{n}}$, respectively. 
The basis states $\{\ket{a}\}$ and $\{\ket{A}\}$ of the subsystems
are taken to be time-independent from now on. 
We can expand an eigenstate in the product basis
$\ket{\Psi_{n}}=\sum_{Aa}\mC^{Aa}_{n}\ket{a}\otimes\ket{A}$,
or vice versa,
$\ket{a}\otimes\ket{A}=\sum_{n}\mD_{Aa}^{n}\ket{\Psi_{n}}$,
where we have the relations
$\sum_{Aa}\mC^{Aa}_{n}\mD^{m}_{Aa}=\delta^{m}_{n}$ and 
$\sum_{n}\mC^{Aa}_{n}\mD^{n}_{Bb}=\delta^{Aa}_{Bb}$.
The expansion coefficients $\mC$ and $\mD$ are time-dependent, 
\begin{equation}
\mC^{Aa}_{n}(t)=\mC^{Aa}_{n}(0)\exp\Big(-\frac{i}{\hbar}\mE_{n}t\Big);
\end{equation}
in any particular case the numerical diagonalization will provide us
with the $\mE_{n}$ and the
coefficients $\mC^{Aa}_{n}(0)$, which make up the eigenvectors,
i.e. $\mC^{Aa}_{n}(0)$ is the $Aa$-component
of eigenvector $n$.

We now can expand an operator $\hat{G}$ in the eigenstates or the
product states:
\begin{equation}
\hat{G}=\sum_{mn}\ket{\Psi_{n}}G^{nm}\bra{\Psi_{m}}=\sum_{ABab}\ket{a}\otimes\ket{A}G^{ABab}(t)\bra{b}\otimes\bra{B},
\end{equation}
with
\be
\label{MatRGt}
G^{ABab}(t)=\sum_{nm}\mC^{Aa}_{n}(0)G^{nm}{\big(\mC^{Bb}_{m}\big)}^{*}(0)\exp\Big(-\frac{i}{\hbar}
\big(\mE_{n}-\mE_{m}\big)t\Big).
\ee
The partial WT with respect to the atomic subsystem is
then
\be
\hat{G}_{w}(R,P,t)=\sum_{ab}\ket{a}G^{ab}_{w}(R,P,t)\bra{b},
\ee
where in turn
\be
\label{MatRGF}
G^{ab}_{w}(R,P,t)=\sum_{AB}\mF_{BA}(R,P)G^{ABab}(t), 
\ee
and
\begin{multline}
\label{MatRFE}
\mF_{BA}(R,P)=\frac{1}{{(2\pi\hbar)}^{n}}\int\exp\Big(\frac{i}{\hbar}P\cdot
s\Big)
\Phi^{*}_{B}\Big(R+\frac{1}{2}s\Big)
\Phi_{A}\Big(R-\frac{1}{2}s\Big){\text{d}}^{n}s.
\end{multline}

Specializing to the particular case of the 2LS and its
actual implementation on a computer, we
introduce dimensionless quantities using the scale factors mentioned
in Sec.~\ref{2LS:sec}.
In particular we obtain the dimensionless wave-function
$\varphi(\xi)=\sqrt{a_{0}}\Phi(a_{0}\xi)$, the
dimensionless version of $\mF$, $F_{nm}(\xi,\eta)
=\hbar\mF_{nm}(a_{0}\xi,b_{0}\eta)$ and dimensionless 
eigenvalues $E_{n}=\mE_{n}/(\hbar\omega)$.
Using Eq.~\eqref{MatRGt}, the factoring Eq.~\eqref{MatRGF}, and
Eq.~\eqref{MatRFE}, all the components of $G^{ab}_{w}$ can be
computed as functions of $R,P,t$ (or their dimensionless
counterparts). 
Note that this does not involve the 
numerical solution of a (potentially partial) differential equation,
so we are not required to 
advance over many small time-steps in order to reach a given value of
$t$. 
Limitations are, however, introduced by
the need to truncate the oscillator basis to a finite number of states. 
As the purpose of the exact approach is to
provide a benchmark for the CEID method, some care has to be taken to
avoid truncation errors here.
First one has to decide how many product and eigenbasis states to use
in expansions like Eq.~\eqref{MatRGt}, say $N$.
Then, the diagonalization has to be carried out using a number of
product basis states $M>N$ such that the $N$ lowest
eigenvalues and corresponding eigenvectors are well converged;
typically, we used $N\approx 120$ and $M\approx 5N$. 
The operator one is considering (in what follows it will be the
density operator) should only have negligible coupling between states 
with index equal to or less than  $N'$ ($N'<N$) to states with index
larger than $N'$. 
These first $N'$ states, which in the diagonalization step are
produced as $M$-component quantities, should not have
significant contributions from product basis components with index
greater than $N$. 
Only if these conditions are met, and the initial conditions for the operator are chosen to involve 
only the first $N'$ states,
can we consider the numerical results for the time-evolution
reliable and an adequate benchmark for CEID.
The quality of the choice of $N$ can therefore only be assessed after
diagonalization.

From the results produced for the purpose of comparison we show the
occupations $N_{a}$, $a\in\{1,2\}$, of the 
electronic levels, and expectation values of position, momentum and
the variance of the position, as functions of time. 
These have been calculated via the WT
$\rho^{ab}_{w}(R,P,t)$ 
of the density operator $\rho$ of the system, by numerical evaluation
of the integral 
\be
N_{a}(t)=\iint\rho^{aa}_{w}(R,P,t)\text{d} R\text{d} P, 
\ee
in case of the occupations, and of
\be\label{exact_observables:eqn}
\langle f(R,P,t) \rangle = \sum_{a=1}^{2}\iint f(R,P,t)\rho^{aa}_{w}(R,P,t)\text{d} R\text{d} P
\ee
for $f(R,P,t)=R$, $P$, $\big(R-\langle R\rangle(t)\big)^{2}$,
respectively, in the rest of the cases.
%
\section{Correlated electron-ion dynamics}\label{CEID:sec}
%
In this section we describe a new formulation of the 
correlated electron-ion dynamics (CEID) while the original formalism
can be found in Ref.~\onlinecite{horsfield04b}.
[See also Ref.~\onlinecite{horsfield05a} and Ref.~\onlinecite{horsfield06a} for
  further details.]
For the sake of simplicity, the new CEID EOM has been derived here
only for the one-dimensional case although multi-dimensional EOM
are also known.
\cite{footnote3}

We start from the well known quantum Liouville equation:
\begin{equation}\label{Liouville:eqn}
\dot{\hat{\rho}} = \frac{1}{i\,\hbar} \, \left[ \hat{H},
  \hat{\rho}\right]\;,
\end{equation}
which is the EOM for the density matrix $\hat{\rho}$ of the
system.
[We use a dot to indicate time-derivative.] 
Unfortunately, a direct integration of Eq.~\eqref{Liouville:eqn} is
exceedingly time-consuming because it scales approximately as the cube
of the Hilbert space dimension which is very large in most cases of
interest.
On the other hand, since atoms are much heavier than electrons,
an expansion of their motion around the classical trajectories
is often justified.
It turns out that this kind of expansion cuts off the quantum
fluctuations of the atomic degrees of freedom 
and so it effectively reduces the Hilbert space dimension.
For instance, simulations of the \emph{semi-classical} limit of 
Eq.~\eqref{Liouville:eqn} 
\cite{martens97} 
have been shown to reproduce --- at least qualitatively ---
the correct non-adiabatic
dynamics of a few interesting test-cases.

More generally, the density matrix $\hat{\rho}$ 
can be partially expanded with respect
to the atomic degrees of freedom by means of a
complete orthonormal system (COS).
\cite{kapral99a}
By using the standard Dirac's bra and ket notation, 
this expansion can be expressed as:
\begin{equation}\label{rho_expansion:eqn}
\hat{\rho} = \sum_{n=0}^{\infty}\sum_{m=0}^{\infty}\,|\phi_n \rangle
\hat{\rho}_{n,m} \langle \phi_m|\;,
\end{equation}
where the functions $\left\{ \phi_n \right\}$ are a COS in the atomic
subspace.
As a consequence, in Eq.~\eqref{rho_expansion:eqn}
the electronic degrees of freedom are
included in 
the \emph{matrix coefficients} $\hat{\rho}_{n,m}$.
A natural choice dictated by the kind of
2LS physics introduced in Sec.~\ref{2LS:sec}
(i.e. confining PES) is to use 
the simple harmonic oscillator (SHO) eigenfunctions as atomic COS.
We stress here that, although these functions are usually centered 
around a classical equilibrium point, any other reference point can be taken
instead (see below).

Since all the observables can be expanded as in
Eq.~\eqref{rho_expansion:eqn},
all the operations involving observables (e.g. averages) can be 
worked out by means of the observable matrix coefficients only.
[In general, the matrix coefficients of the observable $\hat{A}$ are given by:
$\hat{A}_{n,m}= \langle \phi_n| \hat{A}| \phi_m \rangle$.]
For instance, the total energy of the system is given by:
\begin{equation}\label{e_tot:eqn}
E_{tot} = {\rm Tr}\left\{ \hat{H} \hat{\rho} \right\} 
= \sum_{n=0}^{\infty}\sum_{m=0}^{\infty}
{\rm Tr}_e\left\{ \hat{H}_{m,n} \hat{\rho}_{n,m} \right\} \;.
\end{equation}
[The two traces, ${\rm Tr}$ and ${\rm Tr}_e$, apply
to different linear spaces, 
namely the whole Hilbert space and the electronic
subspace: see Sec.~\ref{exact_int:sec}.]
As a further example, 
the EOM for $\hat{\rho}_{n,m}$ are obtained 
by plugging Eq.~\eqref{rho_expansion:eqn} 
into Eq.~\eqref{Liouville:eqn}:
\begin{equation}\label{matrix_EOM:eqn}
\dot{\hat{\rho}}_{n,m} 
= \frac{1}{i\hbar}\,\sum_{k=0}^{\infty}\left[
  \hat{H}_{n,k}\;\hat{\rho}_{k,m} - \hat{\rho}_{n,k}\;\hat{H}_{k,m}
  \right]\;.
\end{equation}
It is easy to prove that $E_{tot}$ is in fact a constant of motion by taking
the time-derivative of Eq.~\eqref{e_tot:eqn} 
and then by using Eq.~\eqref{matrix_EOM:eqn}.

The complete set of EOM,
Eq.~\eqref{matrix_EOM:eqn}, cannot be directly simulated
because it is not finite.
Therefore, we must make an approximation and,
quite naturally, we set to zero (as a matrix) every matrix
coefficient with indices greater than $N$, 
the \emph{CEID order}.
Nevertheless, after this truncation, the EOM are still fully
quantized 
(and expressed by a proper Lie bracket), 
but they are restricted to a smaller Hilbert subspace.

As for the exact scheme described in Sec.~\ref{exact_int:sec}, 
in order to make the classical limit of
Eq.~\eqref{matrix_EOM:eqn} more manifest,
we use a partial Wigner transform (WT) i.e. 
a WT taken only with respect to the atomic degrees of freedom.
Therefore, the partial WT of the operator 
$\hat{A}$, $\hat{A}_w( R, P)$,
is still an operator in the electronic
subspace but it explicitly depends on what are now 
the classical atomic position $R$ and momentum $P$.
In the context of non-adiabatic MD, 
similar partial WT have been already considered
\cite{kapral99a,thorndyke05} 
and they have
been shown to provide correct numerical results.
In order to avoid confusion, we stress that, although
WT seems to map a quantum operator into a classical distribution,
the dynamics remains non-classical because the WT of
the product of two operators is in general not the product of the WT
of the two operators:
\begin{equation}
(\hat{A} \hat{B})_w = \hat{A}_w \star \hat{B}_w \;,
\end{equation}
where $\star$ is the non-commutative \emph{Moyal product}: 
\cite{groenewald46,moyal49} 
\begin{equation}\label{Moyal:eqn}
\star = \exp\left[ \frac{i\,\hbar}{2} \left(
\overleftarrow{\partial}_R\overrightarrow{\partial}_P-
\overleftarrow{\partial}_P\overrightarrow{\partial}_R \right)\right]\;.
\end{equation}
[The arrows indicate the directions in which the derivative operators
  act.]
The WT of any operator can be expanded as in
Eq.~\eqref{rho_expansion:eqn} by using the WT
of the basis operators $|\phi_n\rangle\langle\phi_m|$. 
As a consequence, the matrix coefficients of an operator are the same
both in the original Hilbert space and the in transformed one.

The WT of the Liouville equation can be formally stated as:
\begin{equation}\label{Liouville_WT:eqn}
\dot{\hat{\rho}}_w(t) = \frac{1}{i\,\hbar} \, \left( \hat{H}_w \star
\hat{\rho}_w - \hat{\rho}_w \star \hat{H}_w \right)\;.
\end{equation}
It can be shown that from Eq.~\eqref{Liouville_WT:eqn} the usual
Hamilton-Ehrenfest equations (i.e. the EOM for $\bar{R}={\rm
  Tr}\{\hat{R}\hat{\rho}\}$ and $\bar{P}={\rm
  Tr}\{\hat{P}\hat{\rho}\}$) can be derived: 
\cite{horsfield06a}
\begin{equation}\label{Ehrenfest:eqn}
\left\{
\begin{array}{l}
\dot{\bar{R}} = \bar{P}/M \; ,\\
\dot{\bar{P}} = \bar{F} = -{\rm Tr}\left\{
\left(\frac{\partial \hat{H}_e }{\partial R}\right)
\hat{\rho}
\right\}\;.
\end{array}
\right.
\end{equation}
It is also reasonable to take
the phase-space trajectory $(R(t), P(t))$ as a zero-order  approximation
of the true atomic dynamics if the  quantum fluctuations are not too large.
[The semi-classical limit of the WT is explained more extensively
  in Ref.~\onlinecite{wigner32} (see also Ref.~\onlinecite{groenewald46}).]
This fact suggests that instead of taking the origin as a reference point in the
phase-space $(R,P)$ (i.e a fixed reference
frame), one can take advantage of Eq.~\eqref{Ehrenfest:eqn} and use
$(R(t), P(t))$ as reference (i.e. a mobile reference frame).
After this mobile reference frame transform,
the EOM becomes:
\begin{equation}\label{EOM_MC:eqn}
\dot{\hat{\rho}}_w
= \frac{1}{i\hbar}\,\left[
  \hat{H}_w \star \hat{\rho}_w - \hat{\rho}_w \star \hat{H}_w
  \right] + \left( \frac{\partial \hat{\rho}_w }{\partial \bar{R}} \right) \frac{\bar{P}}{M}
 +\left( \frac{\partial \hat{\rho}_w }{\partial \bar{P}} \right) \bar{F}\;.
\end{equation}
Although it is not apparent from Eq.~\eqref{EOM_MC:eqn}, the EOM can be
still expressed by a proper Lie bracket --- as in
Eq.~\eqref{Liouville_WT:eqn} --- by means of a different
time-translation generator i.e. by a different Hamiltonian.
[Mathematical details can be found in appendix
  \ref{derivation_EOM:sec}.]

At this stage two paths can be followed.
In the first case, the Moyal
product is expanded in $\hbar$ (usually up to first order
\cite{thorndyke05,kapral99a}) 
and a quantum-classical extension of the Liouville equation is
obtained.
However, although this approach is physically appealing, 
one must be aware that the EOM obtained this way cannot be
formulated through a proper Lie bracket,
\cite{caro99,Prezhdo06}
and a generalized non-Hamiltonian bracket should be introduced
instead.
\cite{sergi05}
As a consequence, the evolution of a composite operator, $[AB](t)$,
might be different from the composition of the separated 
evolutions, $A(t)B(t)$,
\cite{caro99,nielsen01}
(because the non-Hamiltonian bracket does not define a proper
derivative),
and the conservation of dynamical symmetries might be a problem
\cite{caro99}
(due to the violation of the Jacobi identity).
It must be also recalled that the difference between the
non-Hamiltonian and Lie brackets is only of order $\mathcal{O}(\hbar)$ 
\cite{nielsen01}
and that the non-Hamiltonian structure arises in a quite natural way for
open systems, whether classical or quantum.
\cite{sergi05}

By following the other route, 
one uses the \emph{exact} expression for the Moyal
product and takes the WT of Eq.~\eqref{rho_expansion:eqn}
as a natural way to truncate $\hat{\rho}_w$.
[We have found a direct expansion in the transformed space 
--- e.g. by using weighted orthogonal polynomials --- 
to cause dangerous instabilities in
the truncated dynamics.]
Therefore, the action of the truncation super-operator ${\rm T}_w$ 
can be defined as follows:
\begin{align}\label{rho_expansion_WT:eqn}
{\rm T}_w\left[ \hat{\rho}_w(R,P,t) \right] &= {\rm T}_w\left[ \hat{\rho}_w(\bar{R} +\Delta R, \bar{P} +\Delta
P,t) \right] \nonumber \\ 
& \equiv \sum_{n=0}^{N}\sum_{m=0}^{N}\,\hat{\rho}_{n,m}(\bar{R},\bar{P},t)
P_{n,m}(\Delta R, \Delta P)\;,
\end{align}
where $P_{n,m}(\Delta R,\Delta P)$ is the WT of $|\phi_n \rangle
\langle \phi_m|$
in the mobile reference frame.
[Properties of these functions are given in appendix
  \ref{CEID_appendix:sec}.]
We opted for this approach because it is still a full quantum scheme ---
but in a truncated Hilbert space --- and the EOM for the density
matrix can still be formulated by means of a proper Lie bracket
(see appendix \ref{derivation_EOM:sec}).

Although an analytical expression of $P_{n,m}$ 
is known, 
\cite{groenewald46} 
it is
far more convenient to state a set of recurrence relations
by taking advantage of the well-known SHO
algebra.
Details can be found in appendix \ref{derivation_EOM:sec};
here we report only the final form
of the CEID EOM for the matrix coefficients $\hat{\rho}_{n,m}$ (in the
mobile reference frame):
\begin{widetext}
\begin{equation}\label{coded_EOM:eqn}
\begin{split}
\dot{\hat{\rho}}_{n,m} &=
 -\frac{b_0^2}{4i\hbar M}\left( \sqrt{(n+2)(n+1)}
  \hat{\rho}_{n+2,m} -(2n+1)\hat{\rho}_{n,m} + \sqrt{n(n-1)}
  \hat{\rho}_{n-2,m} + \right. \\
& \left. -\sqrt{m(m-1)}\hat{\rho}_{n,m-2} +(2m+1)\hat{\rho}_{n,m} 
-\sqrt{(m+2)(m+1)} \hat{\rho}_{n,m+2}\right) + \\
&+\frac{1}{i\hbar} \left[ \hat{H}_e\left( \bar{R} \right),
    \hat{\rho}_{n,m} \right]  -\frac{a_0}{i\hbar} \left( 
  \Delta\hat{F}\left( \bar{R}\right)\sqrt{\frac{n+1}{2}} \hat{\rho}_{n+1,m}
+  \Delta\hat{F}\left( \bar{R}\right)\sqrt{\frac{n}{2}}
\hat{\rho}_{n-1,m} + \right. \\
& \left. -\sqrt{\frac{m}{2}} \hat{\rho}_{n,m-1}\Delta\hat{F}\left( \bar{R}\right)
-\sqrt{\frac{m+1}{2}} \hat{\rho}_{n,m+1}\Delta\hat{F}\left( \bar{R}\right)
\right) +\frac{a_0^2}{4i\hbar}\left( \hat{K}\left( \bar{R} \right)\sqrt{(n+2)(n+1)}
  \hat{\rho}_{n+2,m} + \right. \\
& \left. +\hat{K}\left( \bar{R}
  \right)(2n+1)\hat{\rho}_{n,m} +\hat{K}\left( \bar{R} \right)\sqrt{n(n-1)}
  \hat{\rho}_{n-2,m} -\sqrt{m(m-1)}
  \hat{\rho}_{n,m-2}\hat{K}\left( \bar{R} \right) + \right. \\
&\left. -(2m+1)\hat{\rho}_{n,m}\hat{K}\left( \bar{R} \right) - \sqrt{(m+2)(m+1)}
  \hat{\rho}_{n,m+2}\hat{K}\left( \bar{R} \right)\right)\;,
\end{split}
\end{equation}
\end{widetext}
where $\hat{F} = -\partial \hat{H}_e / \partial R$, $\Delta\hat{F} = 
\hat{F} - \bar{F}$, and $\hat{K} = \partial^2 \hat{H}_e / \partial
R^2$.
[Terms involving higher derivatives of $\hat{H}_e$ should also appear
in Eq.~\eqref{coded_EOM:eqn}, but vanish in this case
since the 2LS Hamiltonian we want to study is quadratic --- see
Eq.~\eqref{elec_H:eqn}.]
We recall that, according to our truncation scheme, one must neglect
in the RHS of Eq.~\eqref{coded_EOM:eqn}
those matrix coefficients  whose indices are greater than the CEID
order.
Those equations, along with Eq.~\eqref{Ehrenfest:eqn}, have been used to
simulate the 2LS dynamics described in Sec.~\ref{2LS:sec}.
In particular, the current implementation uses a second order Runge-Kutta
non-adaptive algorithm to integrate Eq.~\eqref{coded_EOM:eqn} and the
standard velocity-Verlet algorithm to integrate
Eq.~\eqref{Ehrenfest:eqn}.
[A time-step $\Delta t = 10^{-3}/2\pi$ in our natural unit (see
Sec.~\ref{2LS:sec}) has been found to be appropriate for the precision
required by the comparison between CEID an exact approaches reported in
Sec.~\ref{2LS_results:sec}.]
At every integration step, the averaged coordinates, $\bar{R}$ and
$\bar{P}$, are evolved according to Eq.~\eqref{Ehrenfest:eqn} for half a
time-step, then the matrix coefficients are propagated through
Eq.~\eqref{coded_EOM:eqn} for a whole time-step, 
and finally the averaged coordinates are
evolved by another half time-step.
We verified that the accuracy achieved by this kind of symmetric 
Trotter decomposition 
is greater than what is obtained by means of a single evolution
of the averaged coordinates for a whole time-step followed (or
preceded)
by a matrix coefficients propagation for a whole time-step.
Numerical results can be found in Sec.~\ref{2LS_results:sec}.

We now briefly discuss the link between our new
formulation and the original CEID.
%
\subsection{Comparison with former CEID integration schemes}
\label{old_CEID:sec}
%
At variance with the scheme described so far, 
the original formulation of CEID makes use of a completely different
expansion which directly provides EOM for the moments of the density
matrix.
\cite{horsfield05a,horsfield06a}
The most relevant CEID moments are: $\hat{\rho}_e={\rm Tr}_a\left\{ \hat{\rho}
\right\}$, $\hat{\mu}_1={\rm Tr}_a\left\{ \Delta \hat{R} \hat{ \rho}
\right\}$, and $\hat{\lambda}_1={\rm Tr}_a\left\{ \Delta \hat{P} \hat{ \rho}
\right\}$, where ${\rm Tr}_a$ is the partial trace with respect to the
atomic degrees of freedom, $\Delta \hat{R} = \hat{R} - \bar{R}$, and 
$\Delta \hat{P} = \hat{P} - \bar{P}$.
[Higher order moments must be carefully defined
because $\Delta \hat{R}$ and $\Delta \hat{P}$ do not commute.]
On the other hand, analogous objects can be also introduced
in the new formulation:
\begin{equation}\label{mu_nm_def:eqn}
\hat{\mu}_{n,m}(t) = \frac{1}{2\pi\hbar} 
\int {\rm d}R {\rm d}P \, \Delta R^n \Delta P^m \hat{\rho}_w(R,P,t)\;.
\end{equation}
By using the property of the WT, it is easy to find a
link between the new and the original notation:
$\hat{\mu}_{0,0} = \hat{\rho}_e$, $\hat{\mu}_{0,1} =
 \hat{\lambda}_1$, and  $\hat{\mu}_{1,0} = \hat{\mu}_1$.
Similar relations for higher CEID moments can be stated, but some
extra attention must be payed in the derivation
due to the non-trivial commutation relations between positions and
momenta.
It is worth noting that CEID moments provide valuable information 
about the system.
For instance, the quantities
$(\hat{\mu}_{0,0})_{n,n}=(\hat{\rho}_e)_{n,n}$ 
give the probability of observing the system on the $n$-th PES
and the average force (see Eq.~\eqref{Ehrenfest:eqn}) 
can be easily computed from
$\bar{F}= {\rm Tr}\{ \hat{F}(\bar{R}) \hat{\mu}_{0,0}\} 
-{\rm Tr}\{ \hat{K}(\bar{R}) \hat{\mu}_{1,0}\}$.
Higher moments can be used to study electron-ion correlations. 
\cite{horsfield05a}

Moments defined in Eq.~\eqref{mu_nm_def:eqn} 
can be expressed in terms of the matrix coefficients by means of
the following linear transform:
\begin{equation}
\hat{\mu}_{n,m} = \sum_{r,s}\,A^{n,m}_{r,s}\,\hat{\rho}_{r,s}\;,
\end{equation}
where
\begin{equation}
A^{n,m}_{r,s} = \frac{1}{2\pi\hbar}\int {\rm d}R {\rm d}P \Delta R^n
\Delta P^m P_{r,s}(R,P)\;.
\end{equation}
As usual, a set of recurrence relations for $A^{n,m}_{r,s}$ can be
found and --- at least in theory --- 
CEID moments of any order can be computed.
In practice, only the low lying moments are relevant and
here we give a short selection of them:
\begin{widetext}
\begin{subequations}\label{moments:eqn}
\begin{align}
\hat{\mu}_{0,0} &=& \sum_{n=0}^{N} \hat{\rho}_{n,n}\;, \\
\hat{\mu}_{0,1} &=& -i\,b_0\sum_{n=0}^{N} \sqrt{\frac{n}{2}}\left[
  \hat{\rho}_{n,n-1} -  \hat{\rho}_{n-1,n}\right]\;,\\
\hat{\mu}_{1,0} &=& +a_0\sum_{n=0}^{N} \sqrt{\frac{n}{2}}\left[
  \hat{\rho}_{n,n-1} +  \hat{\rho}_{n-1,n}\right]\;,\\
\hat{\mu}_{0,2} &=& -\frac{b_0^2}{2}\sum_{n=0}^{N}\left[ \sqrt{n(n-1)}
  \hat{\rho}_{n,n-2} -(2n+1) \hat{\rho}_{n,n} +  \sqrt{n(n-1)}
  \hat{\rho}_{n-2,n}\right]\;,\\
\hat{\mu}_{1,1} &=& -\frac{i a_0 b_0}{2}\sum_{n=0}^{N}\left[ \sqrt{n(n-1)}
  \hat{\rho}_{n,n-2} -\sqrt{n(n-1)}
  \hat{\rho}_{n-2,n}\right]\;,\\
\hat{\mu}_{2,0} &=& +\frac{a_0^2}{2}\sum_{n=0}^{N}\left[ \sqrt{n(n-1)}
  \hat{\rho}_{n,n-2} +(2n+1) \hat{\rho}_{n,n} +  \sqrt{n(n-1)}
  \hat{\rho}_{n-2,n}\right]\;.
\end{align}
\end{subequations}
\end{widetext}

As anticipated in Sec.~\ref{Intro:sec}, the zero order CEID (i.e. for $N=0$) is
equivalent to the ED: 
\cite{horsfield04b}
\begin{equation}
\dot{\hat{\mu}}_{0,0} = \frac{1}{i\hbar} \left[ \hat{H}_e\left( \bar{R} \right),
    \hat{\mu}_{0,0} \right] \;.
\end{equation}
[We have used Eq.~\eqref{moments:eqn}(a) and Eq.~\eqref{coded_EOM:eqn}.]
We also stress that, in this case, $\hat{\mu}_{0,1} =
\hat{\mu}_{1,0} = \hat{\mu}_{1,1} = 0$ and that both $\hat{\mu}_{0,2}$
and $\hat{\mu}_{2,0}$ are proportional to $\hat{\mu}_{0,0}$.

To clarify the link between the new and the original
formalism, it is helpful to write down the first order ($N=1$) EOM
in terms of the CEID moments.
This can be done by inverting Eqs.~\eqref{moments:eqn}(a-c,f) to express
$\hat{\rho}_{0,0}$, $\hat{\rho}_{0,1}$, $\hat{\rho}_{1,0}$, and
$\hat{\rho}_{1,1}$ as functions of $\hat{\mu}_{0,0}$,
$\hat{\mu}_{0,1}$, $\hat{\mu}_{1,0}$, and
$\hat{\mu}_{2,0}$.
[At the same CEID order $\hat{\mu}_{1,1}=0$ and that $\hat{\mu}_{0,2}=
(b_0^2 / a_0^2)\hat{\mu}_{2,0}$.]
The final result is:
\begin{widetext}
\begin{subequations}\label{first_order:eqn}
\begin{align}
\dot{\hat{\mu}}_{0,0} &=& \frac{1}{i\hbar}\left[ \hat{H}_e(\bar{R}),
  \hat{\mu}_{0,0}\right] -
\frac{1}{i\hbar}\left[ \hat{F}(\bar{R}),
  \hat{\mu}_{1,0}\right]+
\frac{1}{2i\hbar}\left[ \hat{K}(\bar{R}),
  \hat{\mu}_{2,0}\right]\;,\\
\dot{\hat{\mu}}_{0,1} &=& \left\{ \Delta \hat{F}(\bar{R}),
  \hat{\mu}_{0,0} \right\} -\frac{1}{4}\left\{ \hat{K}(\bar{R}),
  \hat{\mu}_{1,0} \right\} +\frac{1}{i\hbar}\left[ \hat{H}_e(\bar{R}),
  \hat{\mu}_{0,1}\right] -\frac{i a_0}{2 b_0}\left[ \hat{K}(\bar{R}),
  \hat{\mu}_{0,1}\right] + \nonumber \\
&&- \frac{1}{a_0^2}\left\{ \Delta \hat{F}(\bar{R}),
  \hat{\mu}_{2,0} \right\} -\frac{b_0^2}{2 a_0^2 M}\hat{\mu}_{1,0}\;, \\
\dot{\hat{\mu}}_{1,0} &=& \frac{1}{2M}\hat{\mu}_{0,1} +  
\frac{1}{i\hbar}\left[ \hat{H}_e(\bar{R}),
  \hat{\mu}_{1,0}\right] -
\frac{i a_0}{2 b_0}\left[ \hat{K}(\bar{R}),
  \hat{\mu}_{1,0}\right] +
\frac{i a_0}{2 b_0} \left[ \hat{F}(\bar{R}),
 \hat{\mu}_{0,0}\right] + \nonumber \\
&&+\frac{a_0^2}{4 b_0^2} \left\{ \hat{K}(\bar{R}),
  \hat{\mu}_{0,1}\right\}\;,\\
\dot{\hat{\mu}}_{2,0} &=& \frac{1}{i\hbar}\left[ \hat{H}_e(\bar{R}),
  \hat{\mu}_{2,0}\right] +
\frac{i a_0}{b_0}\left[ \hat{F}(\bar{R}),
  \hat{\mu}_{1,0}\right]-
\frac{i a_0}{b_0}\left[ \hat{K}(\bar{R}),
  \hat{\mu}_{2,0}\right] +
\frac{3i\hbar a_0^2}{8 b_0^2}\left[ \hat{K}(\bar{R}),
  \hat{\mu}_{0,0}\right] + \nonumber \\
&&+\frac{a_0^2}{2 b_0^2}\left\{ \Delta\hat{F}(\bar{R}),
  \hat{\mu}_{0,1}\right\}\;.
\end{align}
\end{subequations}
\end{widetext}
These equations might be compared with the Eq.~(8) of
Ref.~\onlinecite{horsfield05a} (the mean-field second moment approximation) 
keeping in mind that in that paper the
following \emph{ansatz} as been made: $\hat{\mu}_{2,0} = C^{R,R}
\hat{\mu}_{0,0}$, $\hat{\mu}_{1,1} = C^{R,P}
\hat{\mu}_{0,0}$, and $\hat{\mu}_{0,2} = C^{P,P}
\hat{\mu}_{0,0}$, where $C^{R,R}$, $C^{R,P}$, and $C^{P,P}$ are
time-dependent quantities.
According to our initial conditions (see Sec.~\ref{2LS:sec}), the
initial values of these variables are:
$C_{R,R}(0)=a_0^2/2$, $C_{R,P}(0)=0$ and $C_{P,P}(0)=b_0^2/2$.
In order to find the EOM for $C_{R,R}$, one can trace
Eq.~\eqref{first_order:eqn}(d) and 
it turns out that, to the first order in the coupling constant $f_c$,
$\dot{C}_{R,R}=0$.
Remarkably, by assuming that the matrix $\hat{K}$ is proportional to
the unit matrix and by substituting
$\hat{\mu}_{2,0} = (a_0^2/2)\hat{\mu}_{0,0}$ into
Eq.~\eqref{first_order:eqn}(a-c), we obtain EOM for $\hat{\mu}_{0,0}$,
$\hat{\mu}_{0,1}$, and $\hat{\mu}_{1,0}$ which are equal to the ones
stated in 
Eq.~(8) of Ref.~\onlinecite{horsfield05a} up to the first order in the
coupling constant.
Although there is no reason to believe that this agreement 
must be restricted only to a given set of initial
conditions,
it is not clear yet how it might be proved right 
for higher CEID order or different basis set expansion.
%
\subsection{Energy conservation}\label{energy_conservation:sec}
%
By using the original CEID scheme, it is possible to write the total
energy, Eq.~\eqref{e_tot:eqn}, in terms of the density matrix
moments.
\cite{horsfield04b}
A similar expansion is also obtained by computing Eq.~\eqref{e_tot:eqn}
explicitly.
[The matrix coefficients of the Hamiltonian can be found
in appendix \ref{derivation_EOM:sec}.]
During this computation, 
it is quite useful to distinguish between atomic kinetic energy,
$E_{kin}={\rm Tr}\{ (\hat{P}^2/2M)\hat{\rho}\}$, and atomic
potential energy, $E_{pot}={\rm Tr}\{ \hat{H}_e\hat{\rho}\}$ 
(see Eq.~\eqref{2LS_H:eqn}).
In terms of the CEID moments (see Eq.~\eqref{moments:eqn}), those two
quantities are given by:
\begin{widetext}
\begin{subequations}\label{energies:eqn}
\begin{align}
E_{kin} &=& \frac{\bar{P}^2}{2M}
+\frac{\bar{P}}{M}{\rm Tr}\left\{ \hat{\mu}_{0,1}
\right\} 
+\frac{1}{2M}{\rm Tr}_e\left\{\hat{\mu}_{0,2}
\right\} \;, \\
E_{pot} &=& {\rm Tr}_e\left\{ \hat{H}_e(\bar{R}) \hat{\mu}_{0,0} \right\} 
-{\rm Tr}_e\left\{ \hat{F}(\bar{R}) \hat{\mu}_{1,0} \right\} 
+\frac{1}{2} {\rm Tr}_e\left\{ \hat{K}(\bar{R}) \hat{\mu}_{2,0} \right\}\;, 
\end{align}
\end{subequations}
\end{widetext}
As explained in appendix \ref{correcting_averages:sec}, the CEID
evolution of the bare
averages defined in Eq.~\eqref{energies:eqn} do not give
a conserved (bare) total energy, $E_{tot}= E_{kin} + E_{pot}$, although the
error is negligible for large enough CEID order.
\cite{footnote6}
That is because CEID provides an approximation of the exact
evolution (in the truncated Hilbert space) of the observables' averages
(see Eqs.~\ref{truncated_average:eqn} and
\ref{CEID_average:eqn}).
On the other hand, the \emph{exact} evolution (in the truncated
Hilbert space) of
\emph{every} observable 
can be retrieved starting
from the CEID EOM and then adding a correcting term whose general
analytical expression is reported at the end of appendix
\ref{correcting_averages:sec}

The time-derivatives of the corrections for the bare atomic kinetic
and potential energy --- whose integrals must be
added to Eq.~\eqref{energies:eqn}(a) and Eq.~\eqref{energies:eqn}(b), 
respectively  --- are reported below:
\begin{widetext}
\begin{subequations}\label{res_energies:eqn}
\begin{align}
&\dot{C}_{E_{kin}}^{(N)}
=& +\frac{\bar{P}}{M}(N+1) {\rm
  Tr}\left\{\Delta\hat{F}(\bar{R})\hat{\rho}_{N,N} \right\} +\nonumber \\
&& -\frac{ib_0}{4
  M}(N+1)\sqrt{\frac{N}{2}}{\rm
  Tr}\left\{\Delta\hat{F}(\bar{R})\left(\hat{\rho}_{N,N-1} 
-\hat{\rho}_{N-1,N}\right) \right\} + \nonumber\\
&&+\frac{\bar{P} b_0^2}{4
  a_0 M^2}(N+1)\sqrt{\frac{N}{2}}{\rm Tr}\left\{\hat{\rho}_{N,N-1} +
  \hat{\rho}_{N-1,N} \right\} + \nonumber\\
&&-\frac{\bar{P} a_0}{4M} (N+1) 
\sqrt{\frac{N}{2}}{\rm Tr}\left\{ \hat{K}(\bar{R}) (\hat{\rho}_{N,N-1} +
\hat{\rho}_{N-1,N}) \right\}\;,\\
&\dot{C}_{E_{pot}}^{(N)}
=&-\frac{i a_0^2 \bar{F}}{4
  b_0}(N+1)\sqrt{\frac{N}{2}}{\rm Tr}\left\{\hat{K}(\bar{R}) \left(\hat{\rho}_{N,N-1} -
  \hat{\rho}_{N-1,N} \right) \right\} + \nonumber\\
&&+\frac{i b_0}{4M} (N+1) 
\sqrt{\frac{N}{2}}{\rm Tr}\left\{ \hat{F}(\bar{R}) (\hat{\rho}_{N,N-1} -
  \hat{\rho}_{N-1,N}) \right\}\;.
\end{align}
\end{subequations}
\end{widetext}
%
\section{Comparison between exact integration and CEID}
\label{2LS_results:sec}
%
In this section we present the main results of this work.
They were obtained by means of the two numerical algorithms
described in the previous sections, namely the exact integration
scheme of Sec.~\ref{exact_int:sec} and the CEID scheme 
of Sec.~\ref{CEID:sec}.
We recall that our main goal 
is to attest the convergence of CEID (by increasing its order)
and to verify that the converged results agree with the
exact dynamics of the
2LS geometries introduced in Sec.~\ref{2LS:sec}.
That can be safely done by a direct comparison between 
CEID and exact integration of the time-dependent
Sch\"{o}dinger equation and
this will be the object of Sec.~\ref{CEID_vs_exact:sec} ---
which contains a discussion of the electronic observable dynamics ---
and Sec.~\ref{atomic_observables:sec} ---
which contains a discussion of the atomic 
observable dynamics.

The agreement of CEID with exact integration is clearly 
a fundamental numerical achievement, 
but it does not directly help in the
interpretation of the simulation findings.
Further insights can be obtained by comparing
CEID against analytical
results derived through first-order time-dependent
perturbation theory
(see appendix \ref{time_dep_pert:sec}).
This comparison is reported in
Sec.\ref{CEID_vs_perturbation:sec}.
%
\subsection{Electronic observables}
\label{CEID_vs_exact:sec}
%
Here we present the results of the electronic dynamics.
As initial condition, we always choose the atomic vibrational 
ground-state on the upper PES and than we let the system
to evolve according to either the exact Schr\"{o}dinger evolution or
the CEID equations. 
The WT of the initial $(t=0)$ uncorrelated density matrix is:
$\hat{\rho}_w(\Delta R,\Delta P,0)=P_{0,0}(\Delta R,\Delta
P)\hat{\rho}_e(0)$, where
$P_{0,0}(\Delta R,\Delta P)$, according to the definition given in
Sec.~\ref{CEID:sec},
is the WT (in the mobile reference frame) of the atomic vibrational
ground-state
(which is centered in $\bar{R}=0$ and $\bar{P}=0$) and
\begin{equation}
\hat{\rho}_e(0) =
\left(
\begin{array}{cc}
0 & 0\\
0 & 1
\end{array}
\right)
\end{equation}
describes a pure excited electronic state in the \emph{non-adiabatic}
representation introduced in Sec.~\ref{2LS:sec}.  

The most informative electronic observables are the probabilities to
find the system in the upper or lower electronic state.
Those are obtained as the diagonal entries of the electronic density matrix 
$\hat{\rho}_e= \hat{\mu}_{0,0}$ (see Sec.~\ref{old_CEID:sec}) 
and we shall call them electronic populations.
\begin{figure}[!ht]
\begin{center}
\includegraphics[width=7cm]{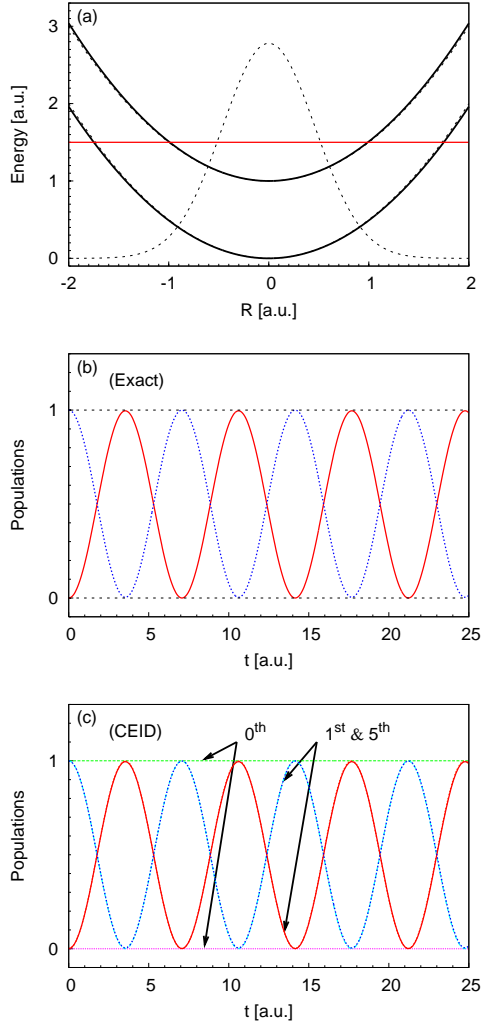}
\end{center}
\caption{
(Color online)
Unshifted 2LS.
(Top) 
Adiabatic energy levels of the electronic
Hamiltonian, $H_e(R)$ (see Eq.~\eqref{elec_H:eqn}) are
plotted (solid lines) against the atomic coordinate, $R$. For the units
employed, see the main text. 
The horizontal line (red online) marks the
the total energy of the system.
A sketch of the initial atomic density distribution is also given
(dashed line). 
(Center)
Electronic populations against time, from exact integration. 
(Bottom)
Electronic populations against time, from CEID. 
Results for CEID order $0$ (i.e. ED), $1$, and $5$ are reported.
}
\label{fig_potential_populations_unshifted:fig}
\end{figure}
In Fig.~\ref{fig_potential_populations_unshifted:fig} we collect the
numerical results for the unshifted case.

A sketch of the PES
is plotted in the first panel. 
[In this particular kind of
2LS, the difference between adiabatic and non-adiabatic surfaces is
negligible.]
In Fig.~\ref{fig_potential_populations_unshifted:fig}(b) 
the exact evolution of electronic populations is reported.
We stress that the 
oscillatory population transfer between the two PES
clearly confirms the crude picture we guessed in Sec.~\ref{2LS:sec}.

In Fig.~\ref{fig_potential_populations_unshifted:fig}(c) the
time-evolution of the electronic populations from CEID
is reported.
Different CEID orders, namely $N=0$, $N=1$, and $N=5$, 
have been considered.
As we expected, the $N=0$ (which is equivalent to ED, see
Sec.~\ref{old_CEID:sec}) does not display any electronic transition.
On the other hand, the outcomes of higher order CEID simulations
present oscillations of the electronic populations
which are in almost perfect agreement 
which the exact integration results.
In particular the first order CEID simulation is well converged
(i.e. there are no visible differences between the $N=1$
and $N=5$ findings).
On the other hand, this is not very surprising;
the unshifted 2LS is the easiest case, since the
symmetries involved ensure that the evolved state is well described by
the simple \emph{ansatz} stated in
Eq.~\eqref{0-order_state:eqn} (see also appendix
\ref{time_dep_pert:sec}).

\begin{figure}[!ht]
\begin{center}
\includegraphics[width=7cm]{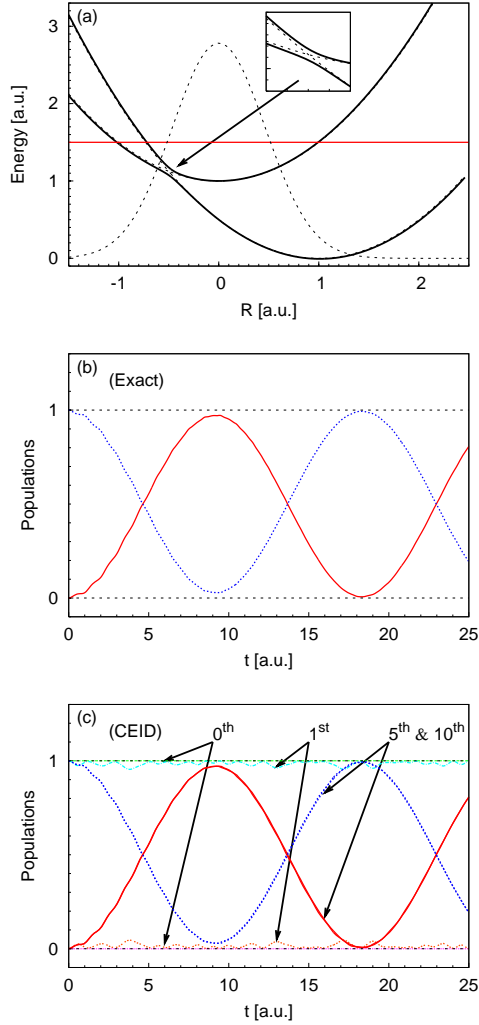}
\end{center}
\caption{
(Color online)
Shifted 2LS.
(Top) 
Adiabatic energy levels of the electronic
Hamiltonian, $H_e(R)$ (see Eq.~\eqref{elec_H:eqn}) are
plotted (solid lines) against the atomic coordinate, $R$.
The avoided crossing is magnified in the inset, where also the
non-adiabatic energies (dashed lines) are reported for comparison. 
For the units employed, see the main text. 
The horizontal line (red online) marks the
the total energy of the system.
A sketch of the initial atomic density distribution is also given
(dashed line). 
(Center)
Electronic populations against time, from exact integration. 
(Bottom)
Electronic populations against time from CEID. 
Results for CEID order $0$  (i.e. ED), $1$, $5$, and
$10$ are reported.
}
\label{fig_potential_populations_shifted:fig}
\end{figure}
In Fig.~\ref{fig_potential_populations_shifted:fig} 
we show the
results for the shifted case following the same scheme employed 
for the previous figure. 
For this kind of 2LS, adiabatic and non-adiabatic PES are
qualitatively different, but
since in Fig.~\ref{fig_potential_populations_shifted:fig}(a)
the difference can be appreciated only close to
the crossing, we provided a magnified plot of that region
in a small inset. 
Once again, almost perfect agreement is seen between a well converged CEID
simulation (here for at least $N=5$) and the exact integration of
the time-dependent Schr\"{o}dinger equation, while at the level of ED
(i.e. $N=0$) the system is stuck in the upper PES.
The fact that a first order CEID simulation is not yet well converged
is not surprising and is a confirmation of 
the general trend predicted in
Sec.~\ref{2LS:sec}:
the larger the surface displacement, $R_0$, 
the higher will be the CEID order
required to obtain a well converged simulation.

We see in
Fig.~\ref{fig_potential_populations_shifted:fig}(b)
that the period of the electronic oscillations is larger for the
shifted case than for the unshifted 2LS.
[A perturbative account of that effect can be found in
appendix \ref{time_dep_pert:sec}.]
Moreover,
in this shifted case the population exchange between the two
PES is not complete:
the minimum of the electronic population on the upper
surface (corresponding to the maximum of the electronic population
on the lower surface) is not exactly
zero (one).
This interesting feature is clearly visible in
Fig.~\ref{fig_potential_populations_shifted:fig}(b)
and is also found in the two well converged CEID simulations
in Fig.~\ref{fig_potential_populations_shifted:fig}(c) so it is not a
numerical feature.
This is instead a non-trivial fingerprint of otherwise elementary
dynamics which is caused by the virtual transitions 
--- a clear quantum effect --- 
between the low lying resonant states and more energetic atomic vibrational
states. 
Further details can be found in Sec.~\ref{CEID_vs_perturbation:sec},
in which we study the dependence of such residual population on
the coupling constant $f_c$. 
%
\subsection{Atomic observables}
\label{atomic_observables:sec}
%
Since the atomic motion is actually non-classical, we expect to find
quantum fluctuations of the atomic observables around their
average values.
We study the time-evolution 
of the average atomic position and momentum, $\bar{R}$ and $\bar{P}$, 
because they provide a sort of effective trajectory
in the phase space which represent an important link 
with classical MD.
Obviously, the concept of trajectory is not well defined in quantum
mechanics and is useful as an approximation
only if the fluctuations are not too large.
So, we also consider the variance of the atomic position,
$\langle \Delta R^2 \rangle$,
in order to test the accuracy of CEID in describing possibly
non-classical atomic dynamics.

\begin{figure}[!ht]
\begin{center}
\includegraphics[width=7cm]{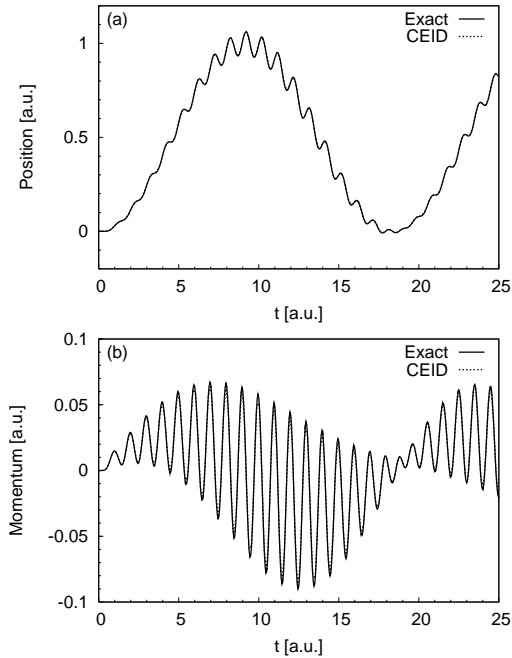}
\end{center}
\caption{
Shifted 2LS.
Plots of the time-evolution of the averaged atomic position, $\bar{R}$, (top) 
and averaged atomic momentum, $\bar{P}$, (bottom). 
Data (almost perfectly superimposed) are taken from exact integration
and well-converged CEID
simulation.
}
\label{fig_R_P:fig}
\end{figure}
We start by reporting results for the average atomic position and
momentum, $\bar{R}$ and $\bar{P}$.
They are evolved by means of the Hamilton-Ehrenfest equations
(see Sec.~\ref{CEID:sec}) according to CEID while in the exact
integration scheme they are obtained by
means of Eq.~\eqref{exact_observables:eqn}.
For the unshifted 2LS, $\bar{R}=0$ and $\bar{P}=0$ for all time 
due to the
inversion symmetry displayed by the system.
On the other hand, the findings reported in Fig.~\ref{fig_R_P:fig} 
for the shifted case once again 
show almost perfect agreement between CEID and exact integration in a
completely non-trivial case.
In particular, CEID not only reproduce the general trend of both
$\bar{R}$ and $\bar{P}$ (large period oscillations),
but also gives the short time scale
details (rapid oscillations).

\begin{figure}[!ht]
\begin{center}
\includegraphics[width=7cm]{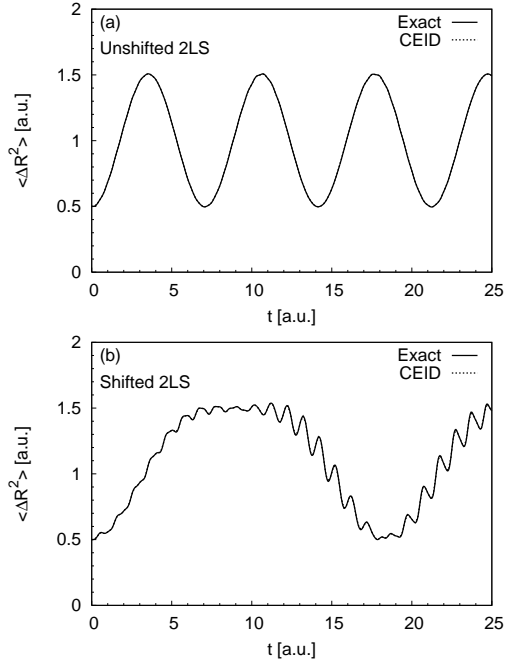}
\end{center}
\caption{
Plots of the time-evolution of the atomic position variance
$\langle \Delta R^2\rangle$, for the unshifted
(top) and shifted (bottom) 2LS.
Data (almost perfectly superimposed) are taken from exact integration
and well-converged CEID
simulation.
}
\label{fig_RR:fig}
\end{figure}
In Fig.~\ref{fig_RR:fig} we report the results for the variance of the
atomic position,  
$\langle \Delta R^2\rangle$, for both the unshifted and
shifted 2LS.
This observable can been obtained 
as the trace of the CEID moment $\hat{\mu}_{2,0}$ defined in
Sec~\ref{old_CEID:sec}.
Once again, almost perfect agreement has been found between a well converged CEID
simulations (here $N=5$ and $N=10$ for the unshifted and shifted
2LS, respectively) and the exact integration of the time-dependent
Schr\"{o}dinger equation.
We stress that such fluctuations are quite significant
and so the atomic dynamics is only poorly approximated by its average
trajectory in the classical phase space.
CEID is working properly even in those highly non-classical cases.

Finally, we have also verified the agreement between CEID and exact
integration for the other entries of the covariance matrix, namely
$\langle \Delta 
P^2\rangle$ and $\langle \Delta R \Delta P\rangle$.
However, numerical findings for those cases are nor reported here
because they are not qualitatively different
from the $\langle \Delta R^2\rangle$ case.
%
\subsection{Comparison between time-dependent perturbation theory and CEID}
\label{CEID_vs_perturbation:sec}
%
In this last section we briefly compare the 
CEID outcomes against 
time-dependent perturbation theory results.
[Mathematical details are collected in appendix
\ref{time_dep_pert:sec}.]

First of all, from a well converged CEID simulation 
(e.g. $N=5$ and $N=10$ for the unshifted and 
shifted case, respectively),
the values of the electronic oscillation frequency
and the residual electronic population can be obtained by
means of a straightforward numerical interpolation.
Then, this procedure can be repeated for the same 2LS geometries, but
different electron-ion coupling constant, $f_c$
(see Eq.~\eqref{2LS_H:eqn}).
It is instructive to study the effect of the atomic motion on the
electronic transitions because it might cause non-classical phenomena,
like quantum interference between different transition paths.
Those effects are usually hard to interpret without a model which
can --- at least qualitatively --- describe the physics involved.
Fortunately, for the kind of model 2LS we considered in this paper, a
simple model can be obtained by means of time-dependent perturbation
theory, whose prediction for the electronic transition frequency and
the residual population are  
summarized in Eq.~\eqref{omega_p:eqn} and Eq.~\eqref{P_res:eqn},
respectively.

\begin{figure}[!ht]
\begin{center}
\includegraphics[width=7cm]{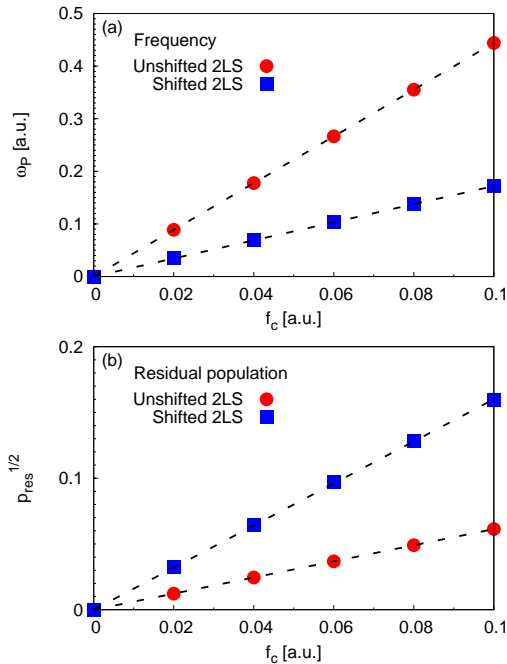}
\end{center}
\caption{
(Color online)
Frequency $\omega_p$ of the electronic transitions
(see Eq.~\eqref{omega_p:eqn}) (top)
and
square root of the residual population $P_{res}^{1/2}$
(see Eq.~\eqref{P_res:eqn}) (bottom)
against the coupling constants, $f_c$, 
for the unshifted and shifted
2LS. 
Linear fits are also showed (dashed lines).
For the units employed, see the main text.
}
\label{fig_fitting_populations:fig}
\end{figure}
In Fig.~\ref{fig_fitting_populations:fig}(a) numerical values 
of electronic population frequencies are reported against several
values of the coupling constant, $f_c$. 
A clear linear trend is manifest in all the 2LS geometries.
Moreover, numerical values are in almost perfect agreement with the
analytical results, Eq.~\eqref{omega_p:eqn}.

In Fig.~\ref{fig_fitting_populations:fig}(b) the residual populations
are plotted against the same coupling constant values.
Although the analytical trend, $P_{res} \simeq \gamma g^2$ (see
Eq.~\eqref{P_res:eqn})
is confirmed, in general is not
easy to give an estimate of the prefactor, $\gamma$.
On the other hand, for the unshifted 2LS case, only one term in
Eq.~\eqref{P_res:eqn} is non-zero
due to the SHO selection rules. 
As a consequence, a numerical estimate can be obtained and it
gives $\gamma \simeq 2.5 \cdot 10^{-1}$, while a direct 
numerical interpolation gives $\gamma \simeq 3.7 \cdot 10^{-1}$.
We stress that such disagreement might depend on the kind of 
approximation we made in order to derive Eq.~\eqref{P_res:eqn} and it
does not effect the general scaling trend of the residual population
with the coupling constant.
%
\section{Discussion and conclusions}\label{conclusions:sec}
%
We have presented a new formulation of correlated 
electron-ion dynamics (CEID).
It is based on a suitable expansion of the quantum
fluctuations around the mean-field atomic trajectories
and its lowest accuracy limit
has been proved to be equivalent 
to the well-known Ehrenfest dynamics (ED). 
This new formulation has been obtained by a
combined use of: 
1) an expansion of the density matrix in terms of
atomic harmonic states centered around the average instantaneous
atomic positions;
2) an \emph{exact} Wigner transform with respect to the
atomic degrees of freedom of the
expanded density matrix. 
The validity of this scheme has been successfully tested
by simulating the non-adiabatic time-evolution 
of a model two level system (2LS).
The accuracy of our simulations is 
determined by a single parameter which 
is related to the order of the density matrix expansion 
and is called the CEID order. 
We then verified that, for all the considered 2LS
 geometries, the exact quantum dynamics --- obtained by exact
 integration of the time-dependent Schr\"{o}dinger equation --- 
is eventually retrieved by increasing the CEID order.
We think that this is a crucial property of our new CEID scheme 
which allows us to estimate the convergence of a numerical simulation 
even when reliable benchmarks are not available.

As for the other proposed CEID schemes,
\cite{horsfield04b,horsfield05a} 
our algorithm only needs 
the Hamiltonian and the initial conditions to start and the subsequent
evolution is computed smoothly, without resorting to any kind of
surface hopping or wave-function spawning.
No \emph{a posteriori} position, velocity, or density
matrix adjustment is needed.
The \emph{exact} evolution (in the truncated Hilbert space) 
of every observable average can be obtained
starting from the CEID EOM by adding a correction term (whose
analytical expression is known) which
is anyhow negligibly small for large CEID order. 
Moreover --- and at variance with other available algorithms
\cite{mueller97,tully98} ---
it works perfectly well within a non-adiabatic 
representation of the electronic PES.
This is desirable because non-adiabatic PES may be smoother
than the adiabatic ones
\cite{ben-nun00}
and also because a costly diagonalization of  
the atomic potential energy $H_e(R)$
at each step is avoided.
All these dynamical properties make our new CEID algorithm a
good candidate for simulating atomic
systems in which quantum coherence 
is relevant.

The advantages of a coherent quantum scheme 
might be relevant even when macroscopic quantum
coherence is not shown.
It is well known that ED --- at variance with other quantum-classical
methods 
\cite{parandekar05a,kapral06}
--- cannot thermalize a
mixed electron-ion system 
(the electronic degrees of freedom are too hot with respect to the
atomic ones.
\cite{parandekar05a})
This failure of the mean-field approximation depends on the absence of
quantum fluctuations which cause spontaneous phonon emission from an
excited electronic state.
[This drawback is apparent also in the 2LS simulations
  considered in this paper (see Sec.~\ref{2LS_results:sec}): the ED is
always stuck in the initial excited state.]
As a consequence, ED does not satisfy microreversibility.
\cite{mueller97,tully98}
On the other hand, a CEID simulation beyond ED can describe
quantum fluctuations and meet the coherence requirements for
microreversibility in a very natural way.
[Once again, see the results of  Sec.~\ref{2LS_results:sec}.]

Although it is not the main concern of this paper, our group is
considering a viable way to 
approach quantum thermalization physics by means of CEID.
A first possibility is given by the spin-boson model
\cite{leggett87} 
in which the bath degrees of freedom are treated explicitly by means
of a collection of many quantum harmonic oscillators.
On a more speculative ground, one can think to implement the
generalization of the Nos\`{e}-Hoover thermostat
introduced in Ref.~\onlinecite{grilli89}.
This scheme is known to fail for ED
\cite{mauri93}
due to the lack of correct quantum back-reaction on the classical bath
variables.
\cite{sergi05}
On the other hand, as we have shown again in this paper, CEID corrects this
ED drawback and it might be better suited for that sort of thermostat.
Moreover, a successful attempt to couple the Nos\`{e}-Hoover
thermostat to the spin-boson model is known in literature
\cite{sergi07a}
and it can provide an interesting test case for future CEID
simulations.

Our CEID algorithm is computationally demanding and
is expected --- in the worst case scenario ---
to scale as  $(N+1)^{2 N_{c} }$, where $N$ is the CEID
order and $N_{c}$
is the number of atomic coordinates.
\cite{footnote5}
Nevertheless, it must be pointed out that the number of relevant
atomic coordinates can be effectively much smaller than $N_{c}$.
\cite{ness99,tamura07}
In this case, one might accelerate a CEID simulation by
allowing for
quantum atomic fluctuations along the relevant directions only.
We also stress that the CEID algorithm is still faster than the
exact integration scheme employed to produce benchmark
calculations in this paper
(see Sec.~\ref{exact_int:sec})
which should scale as $(N+1)^{3 N_{c} }$ since a numerical diagonalization
of the Hamiltonian in the truncated Hilbert space is implied.
We are also considering 
alternative truncations of the Hilbert space
in order to restore the polynomial scaling with atomic degrees of
freedom of the early CEID algorithms.

We see another possible advantage of this CEID scheme over
exact integration: the former expands the quantum fluctuations
around mean-field atomic trajectories, while the latter expands
with respect to a fixed reference frame.
Now, consider a quantum motion in which there are
fluctuations about the mean-field atomic trajectories that are
very tightly confined along a given direction.
With our CEID formulation such fluctuations
can be treated accurately with a low order expansion.
However, schemes that employ
basis functions which are not
centered around the atomic trajectories
could require a very high order expansion to reproduce that confined
behavior if the trajectories are remote from the center of the basis
functions.

Other algorithms, such as molecular dynamics with quantum transitions
or \emph{ab initio} multiple spawning, might have a lower
computational complexity, especially if the region of the 
configuration space where non-adiabatic effects
are relevant is small and crossed only few times during the
time-evolution.
This is the case, for instance, for many chemical reactions in a gaseous
or diluted phase.
On the other hand, we recall here that CEID was explicitly devised to
deal with electron-ion correlations in metals, 
a kind of systems in which the aforementioned algorithms are expected 
to be less efficient.

Needless to say, a reliable algorithm to simulate microscopic 
electro-mechanical effects,
including joule heating, 
will find important application in nanostructure design.
Our CEID algorithm is a good candidate because its accuracy can be
systematically increased by tuning a single parameter that
allows us to approximate the quantum atomic fluctuations in a physically 
transparent way.
Moreover, since quantum coherence is well addressed by CEID, subtle
photo-physical effect like luminescence in conjugated polymers might be
addressed by this method.
Applications of our algorithm to larger atomic systems and
different thermodynamical ensembles are subjects
of ongoing study.
%
%
\begin{acknowledgments}
LS is supported by EPSRC under grant EP/C524381/1 and
MM is supported by EPSRC under grant GR/S80165.
The authors like to thank Tchavdar Todorov for illuminating
suggestions and critically reading of this paper.
MM also thanks A.T. Paxton for helpful comments and
LS acknowledges useful discussions with R. Peixoto Miranda,
D.B. Bowler, P. Delaney, and A.M. Stoneham.
\end{acknowledgments}
%
\appendix
%
\section{Properties of the $P_{n,m}(R,P)$ functions}\label{CEID_appendix:sec}
%
All the properties of the functions $P_{n,m}$ introduced in
Sec.~\ref{CEID:sec} can be obtained by means of the simple harmonic
oscillator (SHO) algebra.

The $P_{n,m}$ are orthonormal:
\begin{equation}\label{orthonormal:eqn}
\langle P_{n',m'}( R, P) , P_{n,m}( R, P) \rangle \equiv
 \frac{1}{2\,\pi\,\hbar}\,\int {\rm d}R
  {\rm d}P \, P_{m',n'}( R, P) \, P_{n,m}( R, P)
  = \delta_{n,n'}\,\delta_{m,m'} \; ,
\end{equation}
and form a complete set in the Wigner space.
[Also note that $\bar{P}_{n,m} = P_{m,n}$.]
The following identities are useful when computing the action of
a canonical operator on $P_{n,m}$:
\begin{widetext}
\begin{subequations}\label{Pnm_useful_identities:eqn}
\begin{align}
R\,\,P_{n,m}(R,P)
&=& +\frac{a_0}{2} \left( \sqrt{\frac{n}{2}}\,P_{n-1,m} +\sqrt{\frac{n+1}{2}}\,P_{n+1,m}
+\sqrt{\frac{m}{2}}\,P_{n,m-1} +\sqrt{\frac{m+1}{2}}\,P_{n,m+1}
\right)\;,\\
\partial_R\,P_{n,m}(R,P)
&=& +\frac{1}{a_0}\left(\sqrt{\frac{n}{2}}\,P_{n-1,m} -\sqrt{\frac{n+1}{2}}\,P_{n+1,m}
+\sqrt{\frac{m}{2}}\,P_{n,m-1} -\sqrt{\frac{m+1}{2}}\,P_{n,m+1}
\right)\;,\\
P\,\,P_{n,m}(R,P)
&=& -\frac{ib_0}{2} \left( \sqrt{\frac{n}{2}}\,P_{n-1,m} -\sqrt{\frac{n+1}{2}}\,P_{n+1,m}
-\sqrt{\frac{m}{2}}\,P_{n,m-1} +\sqrt{\frac{m+1}{2}}\,P_{n,m+1}
\right) \;,\\
\partial_P\,P_{n,m}(R,P)
&=& -\frac{i}{b_0} \left( \sqrt{\frac{n}{2}}\,P_{n-1,m} +\sqrt{\frac{n+1}{2}}\,P_{n+1,m}
-\sqrt{\frac{m}{2}}\,P_{n,m-1} -\sqrt{\frac{m+1}{2}}\,P_{n,m+1}
\right)\;.
\end{align}
\end{subequations}
\end{widetext}
Higher order identities can be recursively 
obtained by using these basic identities.
%
\section{Derivation of the equations of motion}\label{derivation_EOM:sec}
%
In this appendix we show how 
Eq.~\eqref{coded_EOM:eqn} can be derived from Eq.~\eqref{EOM_MC:eqn} by
using the expansion in Eq.~\eqref{rho_expansion_WT:eqn}.
First of all, we want to find the action of the position and momentum
derivatives on the matrix coefficient $\hat{\rho}_{n,m}$.
We formally introduce the operators ${\rm D}_R$ and ${\rm D}_P$
in the following way:
\begin{subequations}
  \begin{align}
    \frac{\partial \hat{\rho}_w }{\partial \bar{R}} &=
    \sum_{n=0}^{\infty}\sum_{m=0}^{\infty}\,{\rm D}_R\left[
    \hat{\rho}_{n,m} \right] 
    P_{n,m}\;, \\
    \frac{\partial \hat{\rho}_w }{\partial \bar{P}} &=
    \sum_{n=0}^{\infty}\sum_{m=0}^{\infty}\,{\rm D}_P\left[
    \hat{\rho}_{n,m} \right] 
    P_{n,m}\;.
  \end{align}
\end{subequations}
Therefore, by means of the identities reported in
Eq.~\eqref{Pnm_useful_identities:eqn}(b,d), we find that:
\begin{subequations}
  \begin{align}
    {\rm D}_R\left[\hat{\rho}_{n,m} \right] &=
    -\frac{1}{a_0} \left( 
    \sqrt{\frac{n}{2}}\,\hat{\rho}_{n-1,m} -\sqrt{\frac{n+1}{2}}\,\hat{\rho}_{n+1,m}
    +\sqrt{\frac{m}{2}}\,\hat{\rho}_{n,m-1} -\sqrt{\frac{m+1}{2}}\,\hat{\rho}_{n,m+1}
    \right) \;,\\
    {\rm D}_P\left[\hat{\rho}_{n,m} \right] &=
    -\frac{i}{b_0} \left( 
    \sqrt{\frac{n}{2}}\,\hat{\rho}_{n-1,m} +\sqrt{\frac{n+1}{2}}\,\hat{\rho}_{n+1,m}
    -\sqrt{\frac{m}{2}}\,\hat{\rho}_{n,m-1} -\sqrt{\frac{m+1}{2}}\,\hat{\rho}_{n,m+1}
    \right) \;. \\
  \end{align}
\end{subequations}
Finally, previous equations can be also written in a more compact form
as follow:
\begin{subequations}
  \begin{align}
    {\rm D}_R\left[\hat{\rho}_{n,m} \right] &=
    \sum_{k=0}^{\infty}\left[
      \hat{D}^{(R)}_{n,k}\hat{\rho}_{k,m} - \hat{\rho}_{n,k}\hat{D}^{(R)}_{k,m}
      \right]\;, \\
	{\rm D}_P\left[\hat{\rho}_{n,m} \right] &=
    \sum_{k=0}^{\infty}\left[
      \hat{D}^{(P)}_{n,k}\hat{\rho}_{k,m} - \hat{\rho}_{n,k}\hat{D}^{(P)}_{k,m}
      \right]\;,
  \end{align}
\end{subequations}
where $ \hat{D}^{(R)}_{n,m} =
-(1/a_0)(\sqrt{n/2}\;\delta_{n,m+1}-\sqrt{m/2}\;\delta_{n+1,m})$ and 
$ \hat{D}^{(P)}_{n,m} =
-(i/b_0)(\sqrt{n/2}\;\delta_{n,m+1}+\sqrt{m/2}\;\delta_{n+1,m})$.
[Here we employed the Kronecker delta.]
As a consequence, the EOM of the matrix coefficients
$\hat{\rho}_{n,m} $ (in the mobile reference frame) can be formally
given by a proper Lie bracket, as in Eq.~\eqref{matrix_EOM:eqn}:
\begin{equation}\label{EOM_MC_Lie:eqn}
\dot{\hat{\rho}}_{n,m} 
= \frac{1}{i\hbar}\,\sum_{k=0}^{\infty}\left[
  \hat{\tilde{H}}_{n,k}\;\hat{\rho}_{k,m} - \hat{\rho}_{n,k}\;\hat{\tilde{H}}_{k,m}
  \right]\;.
\end{equation}
where $\hat{\tilde{H}}_{n,m} = \hat{{H}}_{n,m} + 
i\hbar (\bar{P}/M) \hat{D}^{(R)}_{n,m} + 
i\hbar  \bar{F}    \hat{D}^{(P)}_{n,m}$.
[An operator $\hat{\tilde{H}}$ can be formally defined in the original
Hilbert space taking the anti-Wigner transform of
$\sum_{n=0}^{\infty}\sum_{m=0}^{\infty}\;\hat{\tilde{H}}_{n,m}\;P_{n,m}$.]

Although Eq.~\eqref{EOM_MC_Lie:eqn} is completely equivalent to
Eq.~\eqref{EOM_MC:eqn}, in order to obtain Eq.~\eqref{coded_EOM:eqn}
we still need to compute $\hat{{H}}_{n,m}$.
This can be done by using 
then properties of the $P_{n,m}(R,P)$ functions (see appendix
\ref{CEID_appendix:sec}) and
the following equation:
\begin{equation}\label{matrix_coefficients:eqn}
H_{n,m} =  
\langle
P_{n,m}(\Delta R, \Delta P),
H_w( \Delta R, \Delta P)
\rangle\;.
\end{equation}
For instance, for a quadratic Hamiltonian expanded with respect to the
mobile reference frame:
\begin{equation}
\hat{H}_w = 
\frac{\bar{P}^2}{2M} +\frac{\bar{P} \Delta P}{M} + \frac{\Delta
  P^2}{2M} +\hat{H}_e\left(\bar{R}\right) -\Delta
R\,\hat{F}\left(\bar{R}\right) +\frac{1}{2} \Delta
R^2\,\hat{K}\left(\bar{R}\right)\;,
\end{equation}
one finds that:
\begin{widetext}
\begin{equation}\label{Hnm:eqn}
\begin{split}
\hat{H}_{n,m} &= 
\frac{\bar{P}^2}{2M}\delta_{m,n}
-\frac{i b_0\bar{P}}{M}\left[ \sqrt{\frac{m}{2}} \delta_{m-1,n} -
  \sqrt{\frac{n}{2}} \delta_{m,n-1}\right]
 -\frac{b_0^2}{4M}\left[ \sqrt{m(m-1)} \delta_{m-2,n} + \right.\\
&\left. - (2m+1)\delta_{m,n} +\sqrt{n(n-1)} \delta_{m,n-2}\right]
 +\hat{H}_e\left( \bar{R}\right)\delta_{m,n}
-a_0 \hat{F}\left( \bar{R}\right) \left[
  \sqrt{\frac{m}{2}} \delta_{m-1,n} +\right.\\
&+\left.  \sqrt{\frac{n}{2}} \delta_{m,n-1}\right] 
+\frac{a_0^2}{4} \hat{K}\left( \bar{R}\right) \left[ \sqrt{m(m-1)}
  \delta_{m-2,n} + (2m+1)\delta_{m,n} + \sqrt{n(n-1)}
  \delta_{m,n-2}\right] \;.
\end{split}
\end{equation}
\end{widetext}
[The first three terms account for the the atomic kinetic energy and
  the last three for the atomic potential energy.]
From Eq.~\eqref{Hnm:eqn} one can easily obtain $\hat{\tilde{H}}_{n,m}$
whose expression must be substituted in Eq.~\eqref{EOM_MC_Lie:eqn}
to obtain Eq.~\eqref{coded_EOM:eqn}.
The requirement of neglecting 
in the RHS of Eq.~\eqref{coded_EOM:eqn}
all those matrix coefficients whose indices
are greater than the CEID order  
can be directly enforced by constraining the
summation index $k$ in Eq.~\eqref{EOM_MC_Lie:eqn} to be at most equal to
the CEID order.
%
\section{Origin of the correcting terms to the averages}\label{correcting_averages:sec}
%
The exact evolution of the average of an observable $A$ is given by the
following well-known equation:
\begin{equation}\label{exact_average:eqn}
  \dot{\bar{A}}(t) = \frac{1}{i\hbar}{\rm Tr}\left\{  \hat{A} \left[ \hat{H},
    \hat{\rho}(t) \right] \right\}
  = \frac{1}{i\hbar}{\rm Tr}\left\{  \left[ \hat{A},
    \hat{H} \right] \hat{\rho}(t) \right\}\;.
\end{equation}
This expression is still true in the mobile reference frame since the
extra terms which arise from the implicit time-dependency of $\hat{A}$
and $\hat{\rho}$ cancel out exactly.
The truncated version of Eq.~\eqref{exact_average:eqn} is naturally
given by:
\begin{equation}\label{truncated_average:eqn}
  \dot{\bar{A}}^{(T)}(t) = 
  \frac{1}{i\hbar}{\rm Tr}\left\{  {\rm T}\left[ \left[ \hat{A},
    \hat{H} \right] \right] {\rm T} \left[ \hat{\rho}(t) \right]
  \right\}\;,
\end{equation}
where ${\rm T}$ is the anti-Wigner transform of the truncation
operator ${\rm T}_w$ introduced in Sec.~\ref{CEID:sec}.
[We recall that this operator depends on the CEID order.]
As expected, averages of those observables which commute with the
Hamiltonian are constant of motion.

It turns out that, by using the CEID EOM (see
Eq.~\eqref{coded_EOM:eqn}), the bare dynamics of the averages (in the
mobile reference frame) is given by:
\begin{align}\label{CEID_average:eqn}
  \dot{\bar{A}}^{bare}(t) 
  &= 
  \frac{1}{i\hbar}{\rm Tr}\left\{ \left[ {\rm T} \left[\hat{A} \right],
  {\rm T}\left[  \hat{\tilde{H}} \right] \right] {\rm T} \left[ \hat{\rho}(t) \right]
  \right\} \nonumber \\
  &+ {\rm Tr}\left\{ {\rm T} \left[\frac{\partial \hat{A} }{\partial 
  \bar{R}} \right] {\rm T} \left[ \hat{\rho}(t) \right] \right\}
  \frac{\bar{P}}{M}
  + {\rm Tr}\left\{ {\rm T} \left[\frac{\partial \hat{A} }{\partial 
  \bar{P}} \right] {\rm T} \left[ \hat{\rho}(t) \right] \right\}
  \bar{F} \;.
\end{align}
where we used the modified Hamiltonian
$\hat{\tilde{H}}$ introduced in Sec.~\ref{derivation_EOM:sec}.
We notice that --- unlike the situation for the exact evolution ---
the last two terms in Eq.~\eqref{CEID_average:eqn} do not
cancel out exactly the similar terms coming from the
modified Hamiltonian $\hat{\tilde{H}}$.
For this reason, even if $ [ {\rm T}[\hat{A}],
{\rm T}[  \hat{H} ] ]=0$, $\bar{A}^{bare}$ is not
conserved by the CEID EOM. 
On the other hand,
although the Eqs.~\ref{truncated_average:eqn} and
\ref{CEID_average:eqn} do not provide the same dynamics,
the difference between
$\bar{A}^{(T)}(t)$ and $\bar{A}^{bare}(t)$ --- the correcting term
$C_{\bar{A}}(t)$ --- is expect to be small for
large enough CEID order.
This fact has been verified for the atomic kinetic and
potential energy, whose analytical expressions of the respective
correcting terms are given in Sec.~\ref{energy_conservation:sec}.

Finally, we stress that a general expression for the time-derivative of
the correcting term to the average of the
observable $\hat{A}$ can be derived analytically.
It is given by:
\begin{equation}\label{correcting_term:eqn}
\dot{C}_{\bar{A}}(t) =
-\frac{1}{i\hbar}{\rm Tr} \left\{ {\rm T}\left[ \left[ \hat{O}_{T},
  \hat{\tilde{H}} \right] \hat{A} \right] {\rm T} \left[ \hat{\rho} \right]
\right\} + \mbox{ Hermitian conjugate}\;,
\end{equation}
where $\hat{O}_{T}$ is the Hermitian, idempotent, operator defined by: ${\rm
  T}[\hat{A}] = \hat{O}_{T} \hat{A} \hat{O}_{T} $. 
[The operator $\hat{O}_{T}$ also depends on the CEID order.]
By using this expression, one can implement the correct truncated dynamics of the observable
averages (i.e. Eq.~\eqref{truncated_average:eqn}) 
starting from the CEID EOM and then adding the numerical
integral of the appropriate correcting term.
%
\section{First order time-dependent perturbation theory of a resonant
  2LS}\label{time_dep_pert:sec}
%
We start from the non-interacting limit of Eq.~\eqref{2LS_H:eqn},
i.e. by setting $f_c=0$ in Eq.~\eqref{elec_H:eqn}.
In this easy case, the two PES are dynamically uncoupled.
On each surface 
the atomic configurations can be classified by
using the appropriate SHO basis set. 
We indicate by $|\chi_{n}^{(u)}\rangle$ and $|\chi_{n}^{(l)}\rangle$ 
the $n$-th harmonic excitation of the atomic degrees of freedom
on the upper and lower PES, respectively.
In order to obtain the resonance condition 
between  $|\chi_{0}^{(u)}\rangle$ and
$|\chi_{1}^{(l)}\rangle$ described in
Sec.~\ref{2LS:sec}, we set their energies equal to $E_1 = 3/2 \hbar \omega$.
[Within this setup, the states $|\chi_{n}^{(u)}\rangle$ and 
$|\chi_{n+1}^{(l)}\rangle$ are also degenerate, having 
energy $E_{n+1}= (n+ 3/2)
\hbar \omega$. 
The ground-state is not degenerate and its 
energy is $E_0 = 1/2 \hbar \omega$.]
This degeneracy is lifted
by switching on the interaction ($f_c>0$). 
The hybridization energy is:
$\Delta_{n+1}= \langle \chi_{n}^{(u)}| \hat{H} |
\chi_{n+1}^{(l)}\rangle$.
As a consequence, a system prepared in the state
$|\chi_{0}^{(u)}\rangle$
can decay into the state $|\chi_{1}^{(l)}\rangle$
(see Sec.~\ref{2LS:sec}). 
If the coupling constant is not too large, this process can be treated
perturbatively, with the small parameter
being the \emph{adimensional} coupling constant 
$g=f_c a_0/\hbar \omega$.

We define the unperturbed Hamiltonian $\hat{H}_0$ 
by neglecting all the entries of $\hat{H}$ 
except the diagonal energies and the two off-diagonal
matrix elements coupling $|\chi_{0}^{(u)}\rangle$ and
$|\chi_{1}^{(l)}\rangle$:
\begin{equation}
\hat{H}_0 = \left(
\begin{array}{ccccc}
E_0 &0&0&\cdots& \\
0& E_1 & \Delta_1 &0& \cdots\\
0& \Delta_1 & E_1 &0& \cdots\\
\vdots&0&0& E_2 & \ddots \\
&\vdots&\vdots&\ddots& \ddots  \\
\end{array}
\right)\;.
\end{equation}
The perturbation matrix given by $\hat{V}=\hat{H}-\hat{H_0}$
has zero diagonal entries and does not couple $|\chi_{0}^{(u)}\rangle$ and
$|\chi_{1}^{(l)}\rangle$.

According to our initial condition, the \emph{zero order} solution of the
time-dependent Schr\"{o}dinger equation $i\hbar\partial_t
|\Psi(t)\rangle = \hat{H}|\Psi(t)\rangle$ is given by:
\begin{equation}\label{zero_order_wf:eqn}
|\Psi(t)\rangle \simeq  |\Psi^{(0)}(t)\rangle = 
 e^{\frac{1}{i\hbar}\hat{H}_0\,t}|\chi_{0}^{(u)}\rangle =
 e^{\frac{1}{i\hbar}E_1\,t} \left[ \cos \left(
 \frac{|\Delta_1|t}{\hbar}\right)|\chi_{0}^{(u)} \rangle
-i \sin\left(
 \frac{|\Delta_1|t}{\hbar}\right)|\chi_{1}^{(l)} \rangle\right]\;.
\end{equation}
From Eq.~\eqref{zero_order_wf:eqn}, it is easy to derive the 
dynamics of the upper electronic population:
\begin{equation}
P_u^{(0)}(t) = |\sum_i\; \langle \chi_i^{(u)}| \Psi^{(0)}(t)\rangle
|^2 = \cos^2\left(
 \frac{|\Delta_1|t}{\hbar}\right)\;.
\end{equation}
The correspondent oscillation frequency is given by:
\begin{equation}\label{omega_p:eqn}
\omega_p = 2\frac{|\Delta_1|}{\hbar} \propto  g\,\omega\;
\end{equation}
We stress that at half period ($t = T/2 = \pi/\omega_p$) the upper electronic
population is exactly zero, i.e. at zero order there is no residual
population (see Sec.~\ref{CEID_vs_exact:sec}).

To go beyond the zero order, we use the following transformation: 
$|\Psi(t)\rangle =
e^{\frac{1}{i\hbar}\hat{H}_0\,t}|\phi(t)\rangle$ which leads to an
effective Schr\"{o}dinger equation for $|\phi\rangle$: 
\begin{equation}\label{interaction_picture:eqn}
i\hbar\partial_t
|\phi(t)\rangle = \tilde{V}(t)|\phi(t)\rangle\;,
\end{equation}
where $\tilde{V}(t)=
e^{-\frac{1}{i\hbar}\hat{H}_0\,t}\hat{V
}e^{\frac{1}{i\hbar}\hat{H}_0\,t}$.
This is an interaction picture transformation.
Eq.~\eqref{interaction_picture:eqn} can be solved using the standard
iterative procedure (Volterra's equation)
giving the following \emph{first
  order} approximation of the wave-function:
\begin{equation}\label{first_order_wf:eqn}
|\Psi^{(1)}(t)\rangle = |\Psi^{(0)}(t)\rangle +
 \frac{e^{\frac{1}{i\hbar}\hat{H}_0\,t}}{i\hbar}\int_{0}^{t}{\rm d}\tau\;\tilde{V}(\tau)|
 \chi_0^{(u)}\rangle\;.
\end{equation}
The last term on the RHS of Eq.~\eqref{first_order_wf:eqn} generates
the finite residual population at half period
observed in our numerical experiments.
The following simplified expression holds if $\hbar \omega \gg |\Delta_1|$:
\begin{equation}\label{P_res:eqn}
P_{res} = P_u^{(1)}(T/2) = |\sum_i\; \langle \chi_i^{(u)}| \Psi^{(1)}(T/2)\rangle|^2 \\
= \sum_{i\neq0}\;\frac{|\langle
  \chi_i^{(u)}|\hat{V}|\chi_1^{(l)}\rangle|^2}{(E_{i+1}-E_1)^2}
\propto g^2\;.
\end{equation}
%
%

%
%

\begin{thebibliography}{44}
\bibitem[{foo({\natexlab{a}})}]{footnote4}
\bibinfo{note}{{This approximation may be better understood in terms of a weak
  coupling scheme. This is currently being investigated.}}
\bibitem[{foo({\natexlab{c}})}]{footnote6}
\bibinfo{note}{{Due to the symmetry of the unshifted 2LS, the bare $E_{tot}$ is
  always conserved. For the shifted case, the relative error of the bare
  $E_{tot}$ (i.e. the deviation from perfect conservation) depends on the CEID
  order: It is lesser than $10^{-7}$ for CEID order greater than $10$, but it
  can be as large as $10^{-3}$ for CEID order $5$. By correcting the atomic
  kinetic and potential energy averages as explained in
  Sec.~\ref{energy_conservation:sec}, this relative error becomes lesser than
  $10^{-7}$ at all CEID orders.}}
\bibitem[{foo({\natexlab{d}})}]{footnote5}
\bibinfo{note}{{In $d$ dimensions, $N_{c}$ is $d$ times the number of atoms.
  The dimension of the truncated atomic subspace in which all the atomic
  coordinate fluctuations are expanded up to the same CEID order is $(N+1)^{
  N_{c} }$. As in the rest of the paper, the dimension of the electronic
  subspace is assumed to be fixed.}}
\end{thebibliography}
\end{document}